\documentclass[prb,twocolumn,aps]{revtex4-2}
\usepackage{graphicx}
\usepackage{dcolumn}
\usepackage{bm}
\usepackage{amssymb}
\usepackage{amsmath}
\usepackage{physics}
\usepackage{sublabel}
\usepackage[utf8]{inputenc}
\usepackage{hyperref}
\usepackage[usenames, dvipsnames]{color}
\usepackage[english]{babel}
\usepackage[T1]{fontenc}
\usepackage{bbm}

\usepackage{float}
\usepackage{verbatim}
\usepackage[most]{adjustbox}
\hypersetup{colorlinks = true, linkcolor=blue, citecolor=blue, urlcolor=blue}

  \makeatletter
    \renewcommand\@make@capt@title[2]{%
     \@ifx@empty\float@link{\@firstofone}{\expandafter\href\expandafter{\float@link}}%
      {\textsc{#1}}\@caption@fignum@sep#2\quad}%
    \makeatother

\begin{document}
\title{Exciton magnetic polaron in $\mathbf{CdTe/Cd_{1-x}Mn_{x}Te}$ single semimagnetic quantum ring}
\author{Kalpana Panneerselvam}
\author{Bhaskaran Muralidharan}
\thanks{corresponding author:bm@ee.iitb.ac.in}
\affiliation{Department of Electrical Engineering, Indian Institute of Technology Bombay, Powai, Mumbai-400076, India}

\date{\today}
\begin{abstract}
Magnetically doped nanostructures can significantly enhance the interaction between the band carriers and the dopant atoms. Motivated by the demonstration of the enhanced sp-d exchange interaction in quantum confined structures with the increased stability of the exciton magnetic polaron (EMP), we report the quantitative and qualitative analyses of the EMP formation in $\mathrm{CdTe/Cd_{1-x}Mn_{x}Te}$  diluted magnetic quantum ring (QR). The QR with two different configurations: (i) the non-magnetic ring (CdTe) embedded in the semimagnetic $\mathrm{Cd_{1-x}Mn_{x}Te}$ matrix, and (ii) magnetically non-uniform quantum structures embedded with $\mathrm{Mn^{2+}}$ ions both in the ring and in the barrier regimes, have been investigated for various mole fractions of the Mn dopants. The larger polaron binding energy ($\mathrm{E_{MP}}$) of 23meV is estimated for the 5\% molar Mn contents compared to the other quantum confined systems made of CdMnTe. The magnetic field dependence of the MP energy and the corresponding polaron parameters like exchange field, localization radius of the MP, and the degree of circular polarization induced by the external applied magnetic field at T = 4.2K have been derived. The obtained results are in excellent agreement with the trend of the significant degradation of $\mathrm{E_{MP}}$ in an external magnetic field, and with the contradictory tendencies of $\mathrm{E_{MP}}$ for the QR with configuration (i) and (ii), as reported from the time-integrated measurements based on selective excitation for the quantum systems made of CdMnTe and other DMS materials.
\end{abstract}
\maketitle

\section{Introduction}
\indent The incorporation of paramagnetic dopants into nonmagnetic semiconductors endows the host semiconductors with magnetic functionalities and can achieve the combination of both the magnetism and semiconducting properties in a single material, the so-called diluted magnetic semiconductor (DMS) \cite{furdyna1988diluted,gaj1993relation,kalpana2017donor,rice2017direct}. Such addition of magnetic dopants not only modifies the electrical properties of the semiconductors but can also tune the optical and magnetic properties, enabling the development of optoelectronic and spintronic devices. The giant magneto-optical response of DMS gives rise to exciting features such as giant Zeeman splitting (GZS) \cite{fainblat2016giant, barrows2017excitonic}, giant Faraday rotation \cite{gaj1993giant, panmand2020characterisation}, and the formation of magnetic polaron (MP) \cite{rossin1996magnetic,akimov2017dynamics,kalpana2017bound,kozyrev2021optical}. This originates in the strong sp-d exchange interaction between the spins of the localized charge carriers (electron and hole with s- and p-type wavefunction) and the spins of the 3d-shell magnetic ions. Such interaction is enhanced if the magnetic field is applied and the effective exchange field arises from the polarized magnetic ions in the external field, causing the splitting of the carrier states.\\ 
\indent The counter influence of the exchange field that arises from the localized carrier induces the collective and spontaneous ferromagnetic alignment of the ionic spins within its Bohr orbit, thus generating a net local magnetization even in the absence of any applied magnetic field. The resultant entire spin cloud with a net magnetic moment is known as magnetic polaron, which is a long-standing topic of research. The dynamics of the MP is accompanied by a reduction of the exciton energy, which has been studied using time-resolved spectroscopy and polarization-sensitive techniques \cite{seufert2001dynamical,barman2015time}. The interesting aspects attributed to the MP are the long spin memory times in CdMnTe \cite{gurung2008ultralong} and room temperature ferromagnetic ordering in MnGe quantum dots (QD)\cite{xiu2010room}.\\
\indent The studies in bulk and 2D quantum wells show the key significance of an initial localization of the carriers (exciton) for forming the MP and its stability. The carriers may be bound to an impurity center (donor or acceptor) or by electrostatic and magnetic fluctuations like alloy fluctuations in bulk DMS or interface fluctuations in quantum wells so that the bound and localized magnetic polaron (BMP and LMP) have been studied vigorously in various quantum systems \cite{takeyama1995exciton,zhukov2019optical,elangovan1994bound,mackh1994localized,yakovlev1990first,benoit1993free,mackh1993exciton,poweleit1994observation}. Moreover, the studies show that it is hard to expect the free MP (FMP) stability without initial localization in bulk. In contrast, such an FMP formation could be achieved in semiconductor nanostructures because the confinement reduces the initial localization requirement and thereby enhances the FMP's stability. The extent to which the stability of FMP can be enhanced and controlled via quantum confinement is a significant active area of current research interest and the present work aims at it. The MP and collective magnetic phenomena have been investigated experimentally and theoretically in various $\mathrm{Mn^{2+}}$ doped semiconductor nanostructures, including CdMnTe, CdMnSe epilayers \cite{mackh1994exciton,zayhowski1985picosecond}, and quantum wells \cite{yakovlev1993exciton, yakovlev1995exciton, zhukov2016optical}, and the quantum dots that are grown by molecular beam epitaxy (MBE) \cite{maksimov2000magnetic,gnanasekar2004spin,wojnar2008size,sellers2010robust,anitha2019dynamics} and by the solution-processed colloidal techniques \cite{muckel2017current,barman2020circular,lorenz2020directed}.\\
\begin{figure*}[!htbp]
\begin{adjustbox}{minipage=\linewidth, cfbox=Black 1pt}
       \centering
       \includegraphics[width=0.85\linewidth]{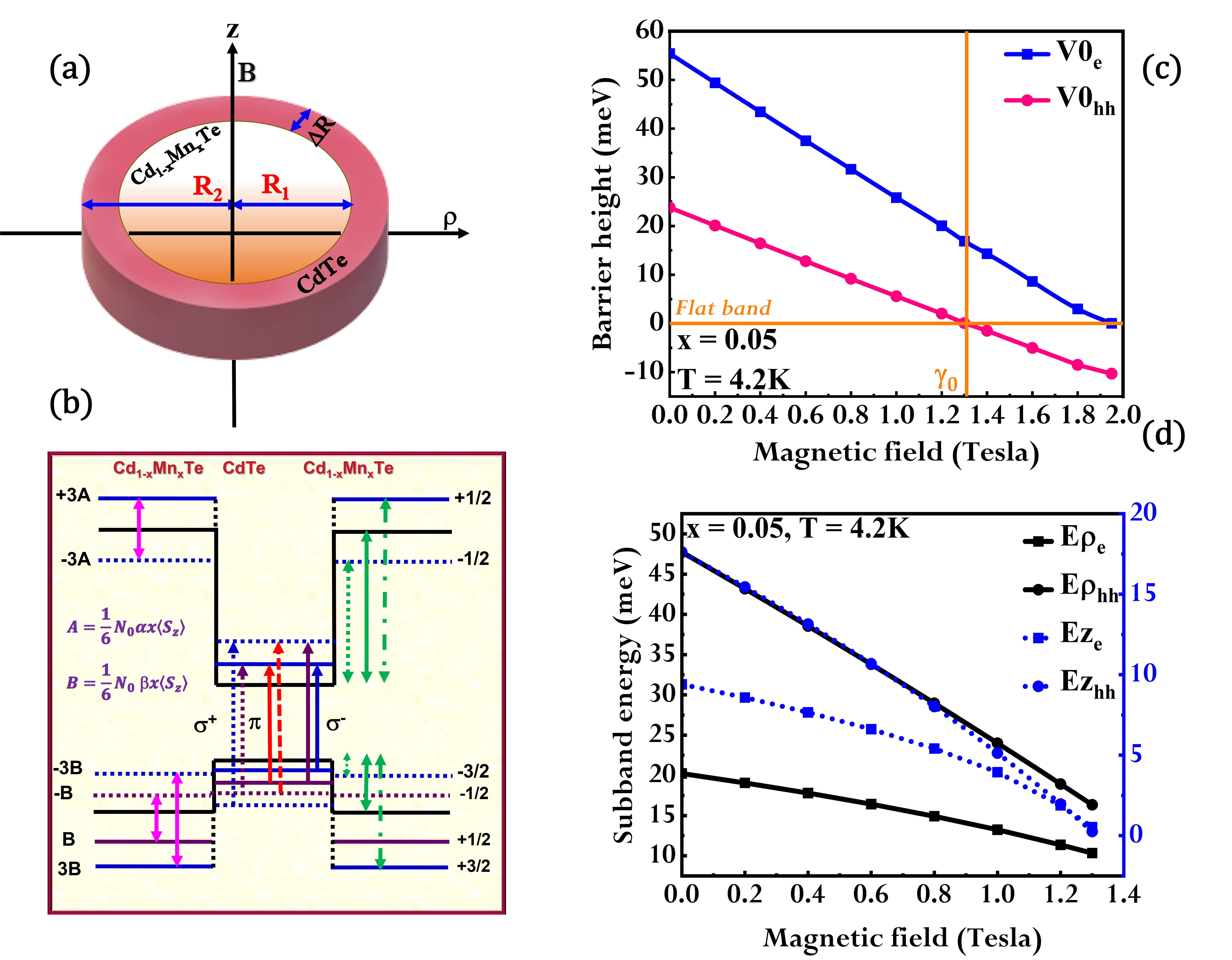}
  \end{adjustbox}
	\caption{Schematic of (a) single $\mathrm{CdTe/Cd_{1-x}Mn_{x}Te}$ quantum ring, (b) Giant Zeeman splitting of the energy levels in nanostructures, (c) Variation of conduction and valence band potential band offset as a function of external magnetic field (d) Variation of subband energies of electron and heavy hole along radial and axial direction of the QR as a function of external magnetic field.}
		\label{Fig.1}
\end{figure*}
\indent Hence, searching for a new EMP regime that novel DMS systems may offer would be exciting. Recent advances in the MBE enable the design of tailored, high-quality nanostructure systems with a wide variety of shapes and complex geometries like doubly connected topological quantum ring (QR) structures. Since QR provides confinement along the radial and axial directions, the formation of exciton magnetic polaron (EMP) can be much more pronounced. However, Mn-containing QRs that deal with the EMP have yet to be developed and to our knowledge it is limited only to BMP \cite{janet2021diluted,sherly2021tuning}. Our present work focuses on a theoretical analysis of the EMP in QR made of well-established II-VI DMS material CdMnTe and to derive the polaron parameters such as polaron energy and the average exchange field. The quantitative assessment of the magnetic field dependence of polaron energy ($\mathrm{E_{MP}}$) and its comparison to GZS for various mole fractions of magnetic dopants in the barrier and/or ring regime has been studied in detail. The ground state exciton energy has been calculated using variational technique in the effective mass approximation and the sp-d exchange interaction has been accounted via mean-field theory.
\section{Theoretical Model }\label{2}
The schematic diagram of a single quantum ring (SQR) is displayed in (Fig. \ref{Fig.1}(a)). The Schrodinger equation and corresponding Hamiltonian for the ground state bound electron-hole pair subjected to a magnetic flux in DMS SQR is written in a dimensionless form, considering the effective Rydberg ($\mathrm{R^{*}}$) as a unit of energy and effective Bohr radius ($\mathrm{a_{B}^{*}}$) as a unit of length, and is given by,
\begin{subequations}
\begin{equation}
\begin{aligned}
&\hat{H}_{ex} \Psi_{ex} = E_{ex} \Psi_{ex}\\ 
\end{aligned}
\label{eqn:1a}
\end{equation}
\begin{equation}
\begin{aligned}
 &\hat{H}_{ex} = -\frac{1}{\rho_{e}^2} \frac{\partial^{2}}{\partial \varphi^{2}} -\frac{1}{\rho_{h}^2} \frac{\partial^{2}}{\partial \varphi^{2}} -\frac{\mu(T)}{m_{e}^{*}(T)} \left(\nabla \rho_{e}^{2} + \nabla z_{e}^{2} \right)\\
   &\quad -\frac{\mu(T)}{m_{h}^{*}(T)} \left(\nabla \rho_{h}^{2} + \nabla z_{h}^{2} \right) + V_{B} (\rho_{e},z_{e}) + V_{B} (\rho_{h},z_{h})\\
 &-\frac{e^{2}}{\epsilon (T) |\Vec{r_{e}}-\Vec{r_{h}}|} +i \, \gamma \frac{m_{h}^{*}-m_{e}^{*}}{m_{h}^{*}+m_{e}^{*}} \frac{\partial}{\partial \varphi} + \frac{\gamma^{2} \rho^{2}}{4}
\end{aligned}
\label{eqn:1b}
\end{equation}
\end{subequations}\\
where, e and h represent the electron and hole, respectively. The strength of the magnetic field is parametrized by $\mathrm{\gamma = \frac{\hbar \omega_{c}}{2 R^{*}}}$, $\mathrm{\omega_{c}}$ is the cyclotron frequency. Since the electron and hole move freely along the annular part of the ring, their motions no longer depend on $\mathrm{\phi_e}$ and $\mathrm{\phi_h}$ separately, but on the relative angular displacement $\mathrm{\phi = \phi_e-\phi_h}$ and it should be treated with the reduced effective mass ‘$\mu$’ of the exciton. Moreover, the material parameters, effective mass, and spatial dielectric constant are considered as temperature-dependent. In the Faraday geometry, the magnetic moments of the ensemble $\mathrm{Mn^{2+}}$ ions with spin angular momentum $\mathrm{S_{Mn}=5/2}$ are subjected to the sp-d exchange interaction with the conduction band electrons of spin $\mathrm{s= 1/2}$ and the heavy hole valence band with angular momentum $\mathrm{J = 3/2}$. This causes the heavy hole exciton splitting into two components with angular momentum +1 and -1, which is composed of $\mathrm{s_{z}=-1/2, \ J_{z}=+3/2, and \ s_{z}=1/2, \ J_{z}=-3/2 }$, respectively as shown in fig. \ref{Fig.1}(b). The sp-d exchange interaction between the electron (hole) and the localized $\mathrm{Mn^{2+}}$ magnetic dopants is denoted by $\mathrm{\hat{H}_{sp-d}}$, and is written as \cite{gaj1993relation, kalpana2017donor, kalpana2019magnetic},
\begin{equation}
\begin{aligned}
\hat{H}_{sp-d} &= -\sum_{i} J(\boldsymbol{r_{e}} - \boldsymbol{R_{i}} ) \boldsymbol{\hat{S}_{i}} \cdot \boldsymbol{\hat{s}_{e}}-\sum_{i} J(\boldsymbol{r_{h}} - \boldsymbol{R_{i}} ) \boldsymbol{\hat{S}_{i}} \cdot \boldsymbol{\hat{s}_{h}}
\end{aligned} 
\label{eqn:2}
\end{equation}\\
‘J’ is the coupling constant for the exchange interaction between the electron (hole) of spin $\mathrm{\hat{s}_{e}}$ ($\mathrm{\hat{s}_{h}}$) located at $\mathrm{\boldsymbol{r_{e}}}$ ($\mathrm{\boldsymbol{r_{h}}}$) and the spin $\mathrm{\hat{S}_{i}}$ of the $\mathrm{Mn^{2+}}$ ions located at sites $\mathrm{\boldsymbol{R_{i}}}$. $\mathrm{V_{B}(\rho_{e,h},z_{e,h})}$ in (\ref{eqn:1b}) is the confining potential of the SQR and is modeled by an abrupt square potential:
\begin{equation}
\begin{aligned}
    V_{B}(\rho_{e,h},z_{e,h}) = \begin{cases} 0 &  R_{1} < \rho_{e,h} \leq R_{2},\\
                                                & \ -d/2 < z_{e,h} \leq +d/2 \\
    V_{0 e,h} & \mathrm{otherwise}
    \end{cases}
    \end{aligned}
    \label{eqn:3}
\end{equation}
 $\mathrm{V_{0 e} = 70\% \Delta E_{g}^{B}}$ , and $\mathrm{V_{0 h} = 30\% \Delta E_{g}^{B}}$ represent the potential band offset formed in the conduction and valence bands, respectively. Tuning of the potential barrier height, $\mathrm{V_{0 e}}$ and $\mathrm{V_{0 h}}$ with the applied field, $\mathrm{B_{z}}$, is possible due to the Zeeman splitting of the band edges (Fig. \ref{Fig.1}(b)) and is written by a formula suggested by K. Navaneethakrishnan et al \cite{jayam2002optical} that satisfactorily fits the experimental Zeeman splitting values available for the $\mathrm{Mn^{2+}}$ compositions x = 0.07, 0.24, and 0.3 with a maximum error of 5\%. Hence, the same formula is adopted here, and the fitting equation is given by \cite{jayam2002optical,gnanasekar2004spin,kalpana2017donor, kalpana2019magnetic},
\begin{equation}
\begin{aligned}
\Delta E_{g}^{B} &= \Delta E_{g}^{0} \, \frac{\eta_{e,h} \, e^{\zeta_{e,h} \, \gamma} - \, 1}{\eta_{e,h} - 1}
\end{aligned}
\label{eqn:4}
\end{equation}
$\mathrm{\Delta E_{g}^{B}}$ and $\mathrm{\Delta E_{g}^{0}}$ denotes the band gap difference between the well CdTe layer and the barrier $\mathrm{Cd_{1-x}Mn_{x}Te}$ layer in the presence and absence of applied magnetic field, respectively. $\mathrm{\eta_{e,h} = e^{\zeta_{e,h} \, \gamma_{0}}}$ is chosen with a fitting parameter $\mathrm{\zeta_{e} (\zeta_{h}) = 0.5 (-0.5)}$, and $\mathrm{\gamma_{0}}$ is a critical magnetic field at which the barrier completely vanishes. The critical magnetic field $\mathrm{\gamma_{0}}$ in Tesla for different magnetic dopant compositions is given for conduction (valence band) as $\mathrm{\gamma_{0}} = A \, e^{nx}$ with A = 0.734 and n = 19.082 (A = - 0.57 and n = 16.706). The Zeeman splitting between the energy levels is well known by the relation \cite{maksimov2000magnetic, akimov2017dynamics},
\begin{equation}
\begin{aligned}
\Delta E_{sp-d}(B_{eff}) &= (\alpha_{exc}-\beta_{exc}) N_{0} \, x \, S_{0} \, \left\langle S_{z}^{Mn} (B_{eff})\right\rangle
\end{aligned}
\label{eqn:5}
\end{equation}
Here, $\mathrm{\alpha_{exc}}$ ($\mathrm{\beta_{exc}}$) is the exchange constant for the conduction band (valence band), which is parametrized for $\mathrm{Cd_{1-x}Mn_{x}Te}$ as $\mathrm{\alpha_{exc} N_{0} = 220meV}$ ($\mathrm{\beta_{exc} N_{0} = -880meV}$) with the atomic concentration of Cd to be $\mathrm{N_{0} = 1.4701 \times 10^{22} cm^{-3}}$. \quad $\mathrm{\left\langle S_{z}^{Mn} (B_{eff})\right\rangle}$ is the thermal average of the spin projection of $\mathrm{Mn^{2+}}$ ions with spin $\mathrm{S_{Mn} = 5/2}$ along the direction of the applied magnetic field B, and is given by the modified Brillouin function, $\mathrm{B_{S}}$, as follows:
\begin{subequations}
\begin{equation}
\begin{aligned}
\left\langle S_{z}^{Mn}(B_{eff})\right\rangle &= \left\langle {\Psi_{ex}^{w}} |{S_{0}(x_{in}) B_{S}(y_{1})}|  {\Psi_{ex}^{w}}\right\rangle + \\
                                            & \quad \left\langle {\Psi_{ex}^{b}} |{S_{0}(x_{out}) B_{S}(y_{2})}|  {\Psi_{ex}^{b}}\right\rangle\\
\end{aligned}
\label{eqn:6a}
\end{equation}
\begin{equation}
\begin{aligned}
B_{S}(y_{j}) &= \frac{2S + 1}{2S} \coth{\frac{2S + 1}{2S}} y_{j} - \frac{1}{2S} \coth{\frac{y_{j}}{2S}} \\ 
&y_{j} = \frac{g_{Mn} \, \mu_{B} \, S_{Mn} \, (B+B_{exc})}{k_{B} \, (T+T_{AF})}
\end{aligned}
\label{eqn:6b}
\end{equation}
\end{subequations}
$\mathrm{g_{Mn} = 2.01}$ is the g-factor of the $\mathrm{Mn^{2+}}$ ion, $\mathrm{\mu_{B}}$ is the Bohr Magneton, $\mathrm{k_{B}}$ is the Boltzmann constant and T is the lattice temperature. For the DMS of arbitrary ‘x’, the antiferromagnetic interactions between the nearest neighbouring $\mathrm{Mn^{2+}}$ ions are included in the calculation through the phenomenological fitting parameters, $\mathrm{S_{0}}$ and $\mathrm{T_{AF}}$, whose numerical values are obtained from the available experimental results \cite{gaj1993relation}. An effective magnetic field is given by $\mathrm{B_{eff}=B+B_{exc}}$. B is the external magnetic field applied along the axial direction of the QR, and $\mathrm{B_{exc}}$ is the exchange field acted upon the $\mathrm{Mn^{2+}}$ ions by the carrier in state which polarizes the Mn spins in its neighbourhood and its relation to the polaron volume is written as \cite{kavokin1999exciton,yakovlev2010magnetic},
\begin{equation}
\begin{aligned}
   &B_{exc}= \frac{1}{\mu_{B} g_{Mn}} \left( \frac{1}{2}\alpha_{exc} s |\Psi_{e}(r)|^{2} + \frac{1}{3} \beta_{exc} J |\Psi_{h}(r)|^{2} \right)\\
   &B_{exc}=\frac{\beta_{exc}}{2 \mu_{B} g_{Mn}}\frac{1}{V_{MP}} ; V_{MP}=\frac{4}{3}\pi r^{3}_{loc}
\end{aligned}
    \label{eqn:7}
\end{equation}\\
In the case of the EMP, the influence of exchange field arises from an electron is neglected and only the heavy hole dominates the polaron energy as the strength of the e-Mn exchange interaction and the overlap of the electron wavefunction with the Mn ions is very much smaller as compared to those of the holes.
The most appropriate trial wavefunction of a ground state exciton is written in a non-separable form due to correlated electron-hole pair as,
\begin{equation}
\begin{aligned}
\Psi_{ex}(r_{e}, r_{h}) &= N_{1s} \, \phi_{e} (\rho_{e}) \, \phi_{h} (\rho_{h}) \, f_{e}(z_{e}) \, f_{h}(z_{h}) e^{-\lambda \, \boldsymbol{r_{eh}}}
\end{aligned}
\label{eqn:8}
\end{equation}\\
where, $\mathrm{\phi_{e,h} (\rho_{e,h})}$, $\mathrm{f_{e,h} (z_{e,h})}$ are the envelope functions along the radial and axial directions, respectively. $\mathrm{e^{-\lambda \, \boldsymbol{r_{eh}}}}$ describes the correlation between the electron and hole which depends mainly on the distance, $\mathrm{r_{eh} = \sqrt{|(\rho_{e}-\rho_{h})|^{2}+|(z_{e}-z_{h})|^{2}}}$ between the two, whereas, $\mathrm{|(\rho_{e}-\rho_{h})|^{2}}$ denotes the projection of the distance between the electron and hole on the plane of the QR and is given by, $\mathrm{|(\rho_{e}-\rho_{h})|^{2}} = (\rho_{e}^{2} + \rho_{h}^{2} -2 \rho_{e} \rho_{h} \cos(\varphi))^{1/2}$.
\begin{subequations}
    \begin{equation}
\begin{aligned}
    \phi (\rho_{e,h}, \varphi_{e,h}) = \begin{cases} \phi _{I} (\rho_{e,h}),  & \rho_{e,h} \leq R_{1} \\
    \phi _{II} (\rho_{e,h}), & R_{1} < \rho_{e,h} \leq R_{2}\\
    \phi _{III} (\rho_{e,h}), & \rho_{e,h} > R_{2}\\
   \end{cases}
    \end{aligned}
    \label{eqn:9a}
\end{equation}
\begin{equation}
\begin{aligned}
    & \phi _{I} (\rho_{e,h}) =  C_{1,e,h} \, I_{0} \left( \beta_{e,h}, \rho_{e,h} \right)   \\
    & \phi _{II} (\rho_{e,h}) = C_{2,e,h} \, J_{0} \left( \alpha_{e,h}, \rho_{e,h} \right) + C_{3,e,h} \, Y_{0} \left( \alpha_{e,h}, \rho_{e,h} \right) \\
    & \phi _{III} (\rho_{e,h}) = C_{4,e,h} \, K_{0} \left( \beta_{e,h}, \rho_{e,h} \right) 
    \end{aligned}
    \label{eqn:9b}
\end{equation}
\begin{equation}
\begin{aligned}
    f (z_{e,h}) = \begin{cases} B_{e,h} \, \exp[k_{e,h} \, z_{e,h}],  &  -\infty < z_{e,h} \leq -d/2 \\
    \cos(\kappa_{e,h} \, z_{e,h}), & -d/2 < z_{e,h} \leq +d/2 \\
    B_{e,h} \, \exp[-k_{e,h} \, z_{e,h}],  &  +d/2 < z_{e,h} \leq +\infty
    \end{cases}
    \end{aligned}
    \label{eqn:9c}
    \end{equation}
\end{subequations}\\
where, $\mathrm{\beta_{e,h} = \frac{m_{b}^{*} (V_{0 e,h}- E_{\rho_{e,h}})}{\hbar^{2}}}$ \; ; \; $\mathrm{\alpha_{e,h} = \frac{m_{w}^{*} E_{\rho_{e,h}}}{\hbar^{2}}}$\\

\; \qquad $\mathrm{k_{e,h} = \frac{m_{b}^{*} (V_{0 e,h}- E_{z_{e,h}})}{\hbar^{2}}}$ \; ; \; $\mathrm{\kappa_{e,h} = \frac{m_{w}^{*} E_{z_{e,h}}}{\hbar^{2}}}$\\

\noindent Here, $\mathrm{N_{1s}}$ is the normalization constant. $\mathrm{C_{1,e,h}}$, $\mathrm{C_{2,e,h}}$, $\mathrm{C_{3,e,h}}$, and $\mathrm{C_{4,e,h}}$ are obtained by choosing proper boundary conditions. $\mathrm{E_{\rho e,h}}$, $\mathrm{E_{z e,h}}$ are the subband energy levels formed due to the radial and axial confinement of the QR. 
The GZS related to that of the exciton is written as \cite{yakovlev2010magnetic,akimov2017dynamics},
\begin{equation}
\begin{aligned}
\Delta E^{hh}_{sp-d}(B_{eff}) = \frac{|\beta_{exc}|}{|\alpha_{exc}|+|\beta_{exc}|} \Delta E_{sp-d}(B_{eff})
\end{aligned}
\label{eqn:10}
\end{equation}
The Photoluminescence(PL)-peak energy in DMS QR corresponding to $\mathrm{\sigma^{+}}$ and $\mathrm{\sigma^{-}}$ can be described by the following expression:
\begin{equation}
\begin{aligned}
\hbar \omega ^{\pm}_{PL} = \hbar \omega_{0} \mp \frac{1}{2}\Delta E^{hh}_{sp-d}(B_{eff})
\end{aligned}
\label{eqn:11}
\end{equation}
The polaron energy is related to the GZS and to the exchange field by\cite{akimov2017dynamics,yakovlev2010magnetic,rice2017direct},
\begin{equation}
\begin{aligned}
&E_{MP}=\frac{1}{2} \Delta E^{hh}_{sp-d}(B_{exc})\\
&\Delta E^{hh}_{sp-d}(B_{exc})=\Delta E^{hh}_{sp-d}(B+B_{exc})-\Delta E^{hh}_{sp-d}(B)
\end{aligned}
\label{eqn:12}
\end{equation}\\
The polaron energy is related to the exchange field by \cite{akimov2017dynamics},
\begin{equation}
\begin{aligned}
&E_{MP}=\frac{1}{2} \gamma_{exc} B_{exc} 
\end{aligned}
\label{eqn:13}
\end{equation}
where, $\mathrm{\gamma_{exc}}$ is the slope of the heavy hole GZS in the limit of weak magnetic fields. Therefore, in the linear approximation, the value of $\mathrm{\gamma_{exc}}$ is related to the slope of the GZS as: $\mathrm{\gamma_{exc}=d \Delta E^{hh}_{sp-d}(B_{eff})/dB|_{B = 0}}$. The slope of the circular polarization degree $\mathrm{\rho_{c}(B)}$ dependence in low magnetic field is $\mathrm{\Theta}=d\rho_{c}(B)/dB|_{B=0}$, which is related to the magnetic fluctuations ($\mathrm{\left\langle M_{f}^{2}\right\rangle}$) \cite{yakovlev2010magnetic}: 
\begin{equation}
\begin{aligned}
&\Theta=\frac{2 \left\langle M_{f}^{2}\right\rangle}{\sqrt{\pi}k_{B}T}; \left\langle M_{f}^{2}\right\rangle=\frac{k_{B}T}{4}\frac{\gamma^{2}_{exc}}{E_{MP}}
\end{aligned}
\label{eqn:14}
\end{equation}
\section{Results and Discussion \label{Res}}
\begin{figure}[!t]
\includegraphics[width=1.01\linewidth]{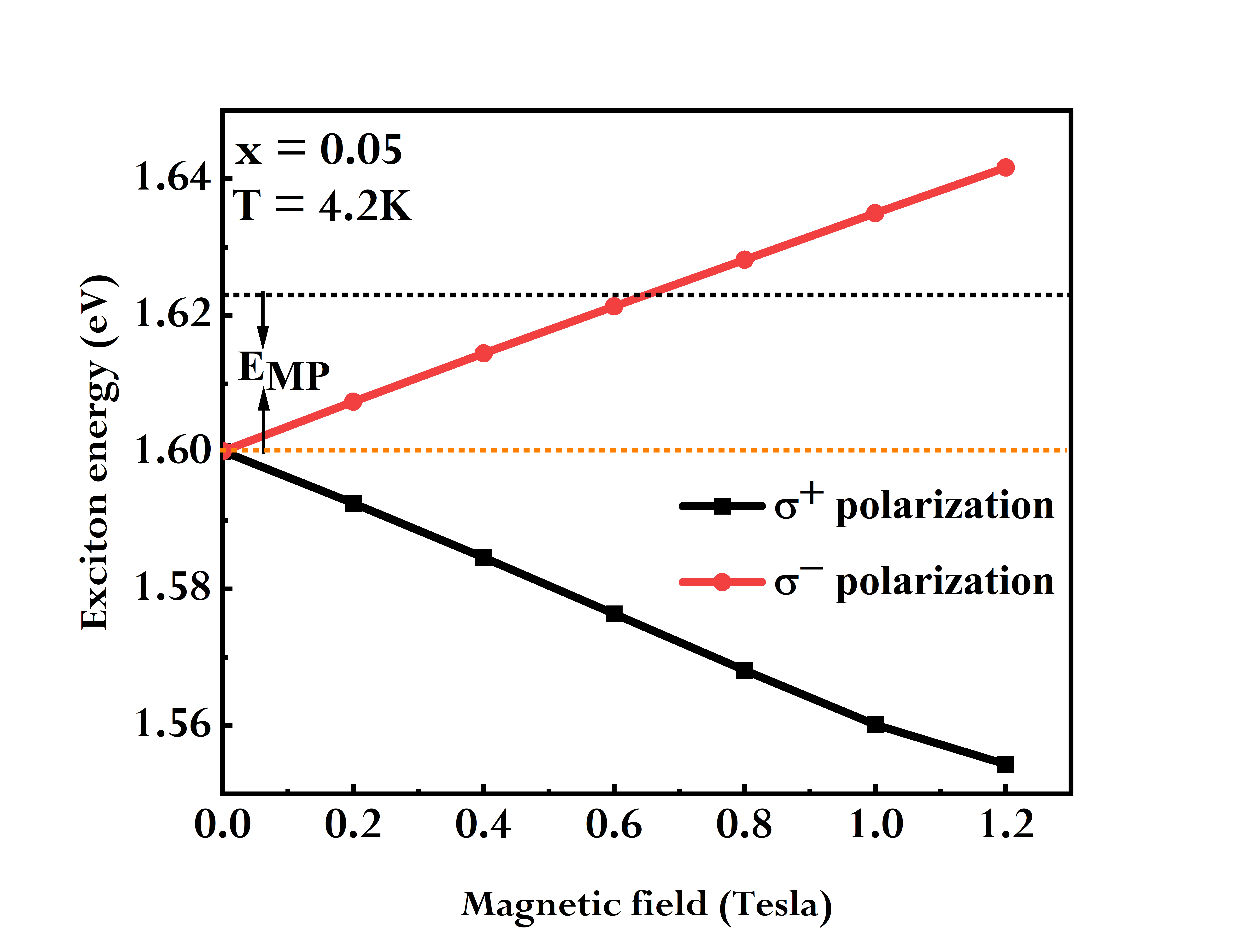}
\caption{Magnetic field dependence of PL peak energy for $\mathrm{\sigma^{+}}$ and $\mathrm{\sigma^{-}}$ magneto-exciton for x = 0.05 at T = 4.2K with the low energy shift of the exciton due to the formation of EMP.}
		\label{Fig.2}
\end{figure}
\begin{figure*}[!htbp]
\begin{adjustbox}{minipage=\linewidth, cfbox=Black 0.9pt}
    \hspace{-1.6em}
    \centering
\includegraphics[width=0.88\linewidth]{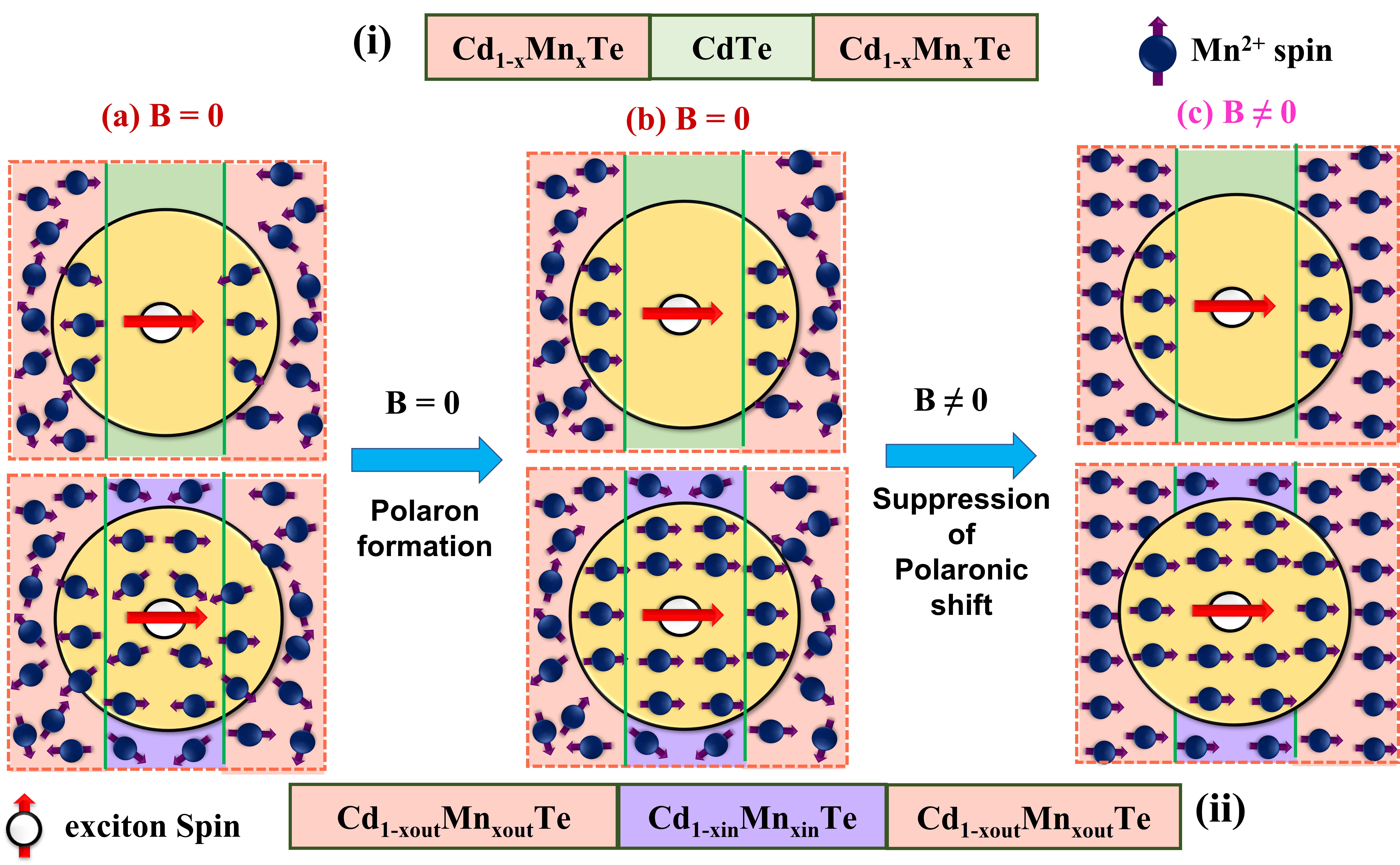}
  \end{adjustbox}
\caption{Schematic representation of polaron formation in (i) non-magnetic QR with magnetic barrier, and (ii) magnetically non-uniform QR structures. Schematic representing the (a) Mn spin orientation before EMP formation, (b) EMP formation in the absence of magnetic field, and (c) Polaron suppression in external magnetic field.}
\label{Fig.3}
\end{figure*}
\indent In fig. \ref{Fig.2}, the magnetic field dependence of the photoluminescence (PL) peak energy corresponding to $\mathrm{\sigma^{+}}$ and $\mathrm{\sigma^{-}}$ polarization is shown for $\mathrm{CdTe/Cd_{1-0.05}Mn_{0.05}Te}$ QR maintained at T = 4.2K. The figure shows that at B = 0, the PL is unpolarized, i.e., the PL peak energies of $\mathrm{\sigma^{\pm}}$ magneto-exciton are degenerate. However, the applied magnetic field breaks the degeneracy and causes the PL to split into left ($\mathrm{\sigma^{-}}$) and right ($\mathrm{\sigma^{+}}$) circularly polarized. This is indicated in fig. \ref{Fig.2} by a monotonic shift of the exciton energy towards low and high energies about the zero-field value, and the PL gets resolved into two branches of exciton doublet corresponding to $\mathrm{\sigma^{-}}$ and $\mathrm{\sigma^{+}}$ polarization, respectively. These two contradicting trends of the rise and drop in the PL peak energy, as B is increased, are the telltale hallmark of any semiconductor heterostructures made of DMS that have been previously observed in CdMnTe epilayers \cite{mackh1994localized}, DMS quantum wells \cite{mackh1993exciton, akimov2017dynamics}, and quantum dots \cite{maksimov2000magnetic, rice2017direct}. A quick recollection of the following facts can help us to understand these trends in the PL peak energy with B: The spin-up and spin-down carriers experience different degrees of confinement inside the QR under the applied magnetic field. This is because of the dynamics pertaining to the band edges of the DMS barrier layers for both the spin-up and spin-down carriers are different. The band edges move closer to each other for the spin-down carriers, which is responsible for decreasing the potential band offset, and vice versa for the spin-up carriers. Hence, the applied B tremendously reduces the potential band offset formed in the valence and conduction band ($\mathrm{V0_{hh}}$ and $\mathrm{V0_{e}}$) for the spin-down carriers. Once it reaches the critical value ($\gamma_{0}$), the valence band attains a flat band situation. However, $\mathrm{V0_{e}}$ is not very sensitive to the applied field, so the presence of $\mathrm{V0_{e}}$ could be observed for the critical field value at which $\mathrm{V0_{hh}}$ vanishes. This is because the exchange interaction that the $\mathrm{Mn^{2+}}$ ions have with the holes in the valence band ($\mathrm{|\alpha_{exc} N_{0}|=220meV}$) is always larger than with the electrons in the conduction band ($\mathrm{|\beta_{exc} N_{0}|=880meV}$) which leads to a larger splitting in the former rather than in the latter. Hence, the value of $\mathrm{\gamma_{0}}$ at which the potential barrier is completely suppressed is always different for the conduction and valence band. Therefore, the magnetic tuning of the potential barrier leads to the tuning of the subband energy levels of both the electron and hole in a unique way, affecting the excitonic bound states. To substantiate these facts, the variation of $\mathrm{V0_{hh}}$ and $\mathrm{V0_{e}}$ with B and corresponding subband energy levels have been plotted in fig. \ref{Fig.1}(c) and \ref{Fig. 1}(d) for the $\mathrm{Mn^{2+}}$ composition of x = 0.05 at T = 4.2K. It is apparent from fig. \ref{Fig.1}(c) that a magnetic field strength of B = 1.3T is required for complete suppression of $\mathrm{V0_{hh}}$, whereas B = 1.95T is required for the same in the conduction band. Similarly, a rapid decrease of subband energy levels of the heavy hole could be seen as compared to the electron, which reflects in the excitonic energy states.\\ 
\begin{figure*}[!htbp]
\begin{adjustbox}{minipage=\linewidth, cfbox=Black 0.9pt}
\centering
\includegraphics[width=0.9\linewidth]{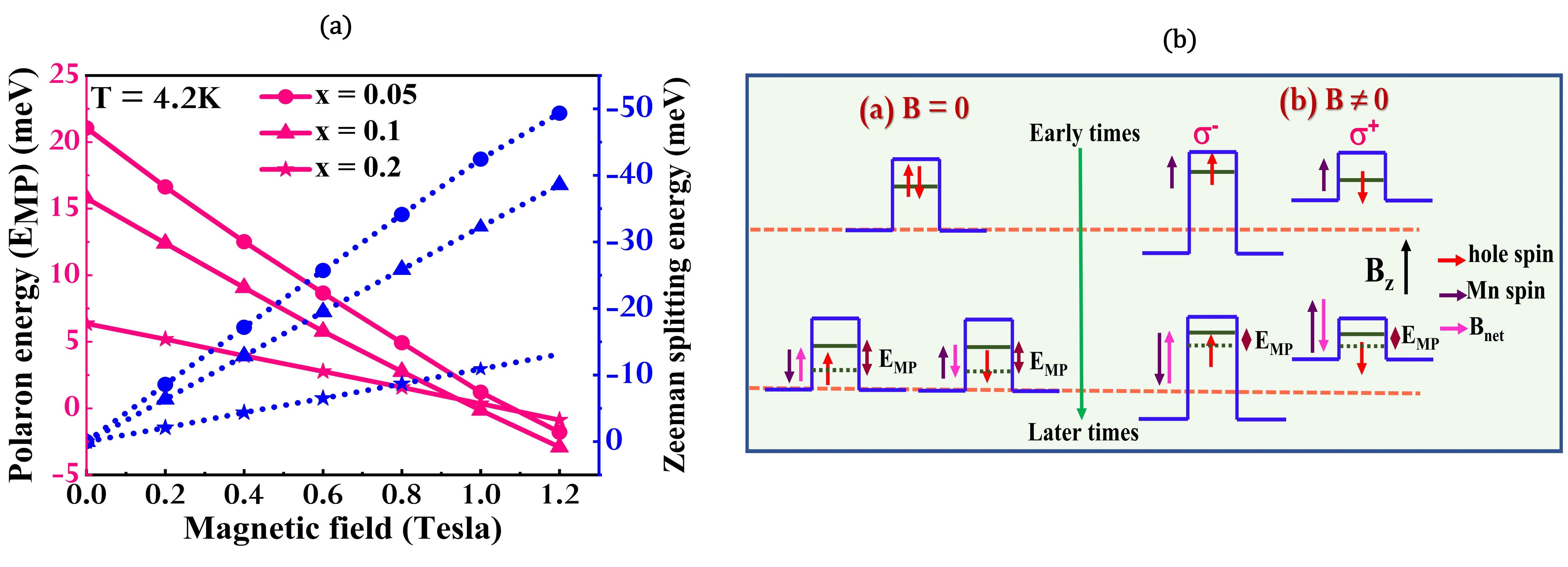}
  \end{adjustbox}
\caption{(a) Magnetic field dependence of polaron energy and GZS of the energy levels for the QR with increasing molar contents of x = 0.05, 0.1, and 0.2 in the barrier layer. (b) A model explaining the spin orientation of the Mn spin sub lattice and the exciton spin during EMP formation.}
		\label{Fig.4}
\end{figure*}
\indent The interesting part of the energy spectrum depicted in fig. \ref{Fig.2} lies in the lowering of the exciton energy from its excitation value in the absence of a magnetic field due to the formation of MP. When the exciton is selectively excited into its localized band states, it starts to have an exchange interaction with all the magnetic moments of the $\mathrm{Mn^{2+}}$ ions that fall into its effective Bohr radius by exerting an exchange field ($\mathrm{B_{exc}}$) which amounts to several Tesla. This exchange field exerted by the exciton ferromagnetically polarizes all the moments in its orbit, and the resultant spin cloud with a finite value of the net magnetic moment is known as an EMP. These polarized $\mathrm{Mn^{2+}}$ ions act back on the exciton and lower its energy, so the recombination will take place from the lowest final state instead from the initial localized band state. This fact is well followed by the exciton confined in DMS QR, as one can notice from fig. \ref{Fig.2} that there is a low energy shift in the exciton energy from 1.623eV to 1.60eV. This energy difference of 23.4meV is known as the energy that is lost by the exciton due to the formation of MP and is represented by the EMP binding energy ( $\mathrm{E_{MP}}$). This evidences further exciton localization in addition to the confinement provided by the QR heterostructure. Such a formation of EMP has a great impact on the exciton mobility, which can be determined experimentally by measuring “exciton mobility edge,” representing a specific energy below which the exciton motion is completely arrested, and the spectral diffusion due to the phonon-assisted tunnelling does not occur during the exciton lifetime \cite{yakovlev2010magnetic, akimov2017dynamics}. The manifestation of the magnetic polaron can be elucidated by the time-resolved resonant photoluminescence experiment in which the large temporal red-shift of the PL peak energy from the selective excitation energy in the absence of magnetic field and the suppression of this shift with the magnetic field and temperature confirms the magnetic polaron formation \cite{rice2017direct, seufert2001dynamical, barman2015time}. But the present work aims to analyze the formation of EMP in DMS QR through theoretical calculations, which leads to the motivation for experimental verifications.\\
\begin{figure*}[!htbp]
\begin{adjustbox}{minipage=\linewidth, cfbox=Black 1pt}
    \hspace{-1.6em}
    \centering
\includegraphics[width=0.802\linewidth]{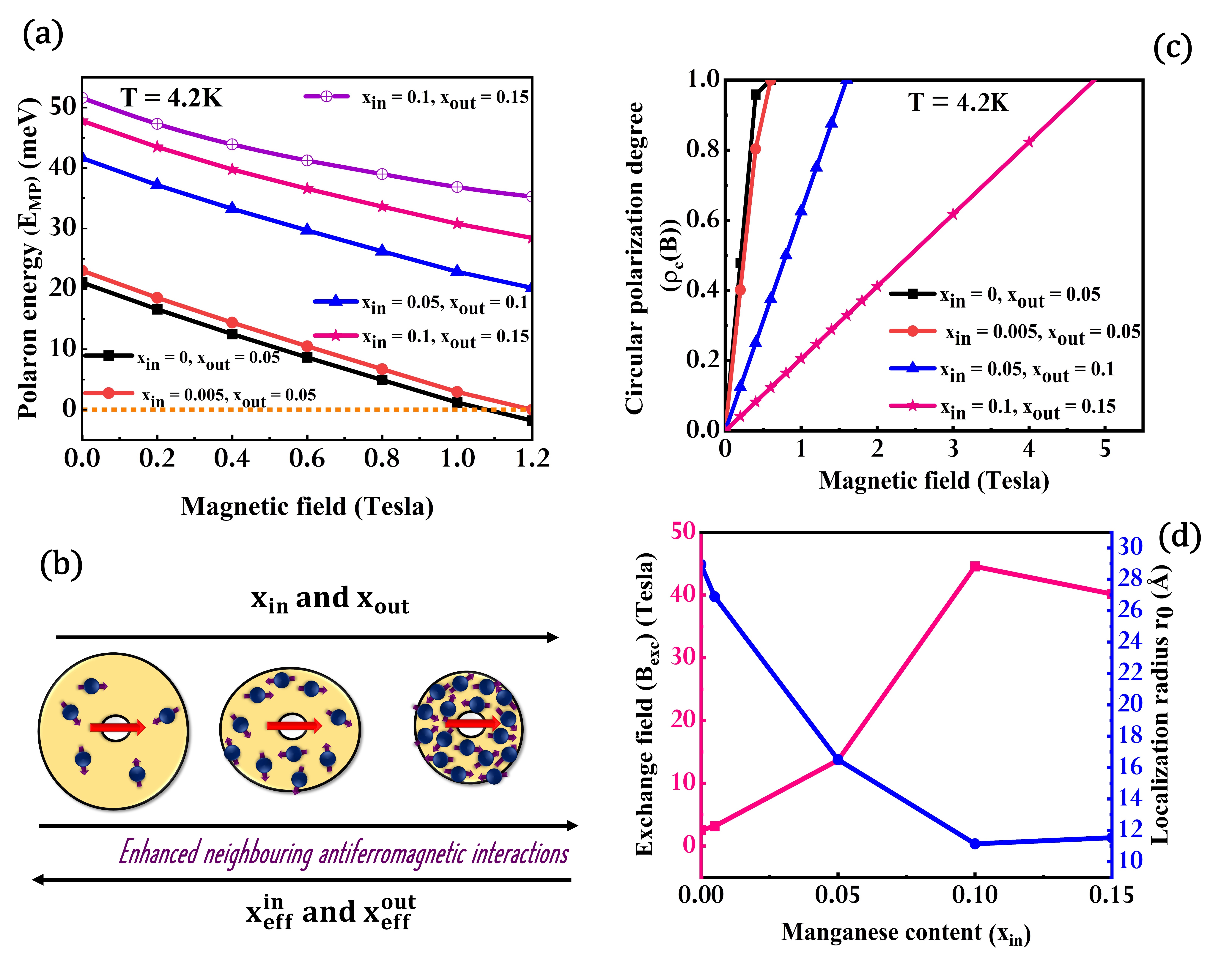}
  \end{adjustbox}
\caption{(a) Magnetic field dependence of polaron energy for the simultaneous increase of Mn contents both in ring and in the barrier regimes at T = 4.2K. (b) Schematic representation of the reduction of localization radius of the exciton with the increase of Mn content inside the ring region. (c) Variation of circular polarization degree as a function of magnetic field for various values of $\mathrm{x_{in}}$ and $\mathrm{x_{out}}$. (d) Effective exchange field ($\mathrm{B_{exc}}$) and the localization radius of the exciton as a function of Mn contents doped in ring regime.}
		\label{Fig.5}
\end{figure*}
\indent The calculations have been performed for two different configurations: (i) QR made of non-magnetic material (CdTe) surrounded by magnetic barriers ($\mathrm{Cd_{1-x}Mn_{x}Te}$) doped with various concentrations of the magnetic ions (x = 0.05, 0.1, and 0.2), and (ii) magnetically non-uniform QR structures like both the ring ($\mathrm{Cd_{1-x_{in}}Mn_{x_{in}}Te}$) and the barriers ($\mathrm{Cd_{1-x_{out}}Mn_{x_{out}}Te}$) are doped with various mole fractions of $\mathrm{Mn^{2+}}$ ions in such a way that the difference between the concentrations in both sides of the interfaces always yields the same concentration x = 0.05 so that the potential band offset is maintained to be same. The upper and lower panel in fig. \ref{Fig.3} denote the configuration (i) and (ii), respectively. Fig. \ref{Fig.3}(a) corresponds to the situation before the MP formation in the absence of a magnetic field where the magnetic ions in and outside of the exciton Bohr orbit are randomly oriented, ensuring there is no net magnetic moment. But at later times, the formation of MP occurs by polarizing the magnetic ions inside its orbit, and thereby the EMP gains the net magnetic moment,$\mathrm{M_{p}}$, as shown in fig. \ref{Fig.3}(b). The entire scenario is modified by the application of an external magnetic field since it aligns the randomly oriented magnetic moments outside the orbit, which suppresses the formation of EMP (fig. \ref{Fig.3}(c)). To corroborate these facts, the calculation of polaron binding energy has been computed, and the results are discussed in the subsequent sections.\\
\indent Turning to the configuration (i), which is frequently encountered experimentally, the magnetic field dependence of the variation of polaron energy and the Zeeman splitting of the energy levels at T = 4.2K for x = 0.05, 0.1, and 0.2 in the barrier DMS layers have been plotted in fig. \ref{Fig.4}(a). It is noted from figure that irrespective of the concentration of the doped magnetic ions, the polaron energy drastically decreases with B, which is a conventional trend observed in various DMS epilayers and heterostructures\cite{mackh1994exciton, yakovlev1992two, barman2015time}. The polaron energy for x = 0.05 is reduced from 23.4eV at B = 0 to half of its value at B = 0.6T. This suppression of the polaron energy in the magnetic field is attributed to the large Zeeman splitting of the energy levels, which arises from the reaction of the carriers to the total average spin alignment in the global Mn spin system. In contrast, the polaron energy is only affected by the local spins of the $\mathrm{Mn^{2+}}$ ions residing in the polaron orbit. The Zeeman splitting of the energy levels in a given magnetic field measures the degree of spin alignment at that particular magnetic field. Hence, the polaron suppression ensures the degree of spin alignment outside and inside the Bohr orbit approaches each other, which could be seen from the complete suppression of polaron formation in fig. \ref{Fig.4}(a) when the Zeeman splitting is about to reach its saturation value. The internal exchange field ($\mathrm{B_{exc}}$) acted upon the local spins by the hole is calculated using Eq (\ref{eqn:7}), and it is found that the field of $\mathrm{B_{exc}=2.54T}$ is responsible for the total alignment of the local spins inside the Bohr orbit for x = 0.05, which is always larger than the external field required to align the global Mn spin system. Since $\mathrm{B_{exc}}$ is inversely proportional to the exciton localization volume ($\mathrm{V_{MP}}$), the latter is found to be $\mathrm{V_{MP}=1495N_{0}^{-1}}$.\\
\indent Moreover, an increase of Mn content beyond 5\% does not favour the polaron formation, as noted from the reduced value of the polaron energy for x = 0.1 and 0.2 and from the difference in the slopes for all Mn contents in low magnetic fields. When the DMS QR is formed with configuration (i), the interaction between the magnetic ions embedded in the barrier layers with the total wavefunction of the carriers localized in the ring is impossible. However, it could interact with the part of the carrier wavefunction because of its protrusion through the interface. Once the concentration increases, the antiferromagnetic interaction between the nearest neighbours also increases, which has been incorporated in the calculation through $\mathrm{S_{0}}$ and $\mathrm{T_{0}}$ parameters and leads to the hardening of the magnetic system. This cancels out the nearest paramagnetic spins and thereby reduces the number of isolated spins contributing to the spin's thermal average, which minimizes the molar concentration 'x' to its effective value '$\mathrm{x_{eff}}$'. For $\mathrm{Cd_{1-x}Mn_{x}Te}$, experimental results \cite{gaj1993relation, lee1996magneto, jain1991diluted} show that the value of $\mathrm{x_{eff}}$ increases until x reaches 0.05 and beyond which starts decreasing. Apart from this, the incorporation of large Mn contents in the barrier does not affect the effective Bohr radius of the carrier. Therefore, the number of effective Mn ions with which the penetrated carrier wavefunction interacts diminishes, reflecting in the polaron energy estimation.\\
\indent Based on the model reported in \cite{kochereskho1994light, sellers2010robust}, further understanding of the orientation of the Mn and the carrier spins in the absence and presence of a magnetic field can be explained using the schematic representation in fig. \ref{Fig.4}(b). The upper and lower panel in figure explains the spin orientation of the entire system at initial and final times, respectively. The direction of the magnetic field is considered to be perpendicular to the plane of the QR. In the absence of a magnetic field, since the localized spins are randomly oriented, there is no net magnetic moment of the Mn spin system, and thereby the energy levels corresponding to the spin-up and spin-down carriers degenerate. The direction of the exchange field $\mathrm{B_{exc}}$ and the net magnetic field $\mathrm{B_{net}=B_{exc}}$ always prefers the orientation of the hole spin. As time evolved, MP formation occurs by polarizing all the localized magnetic moments antiparallel to the hole's spin due to h-Mn exchange interaction, which reduces the carrier energy with a shift of $\mathrm{\Delta E}$ called the polaronic energy shift, as shown in fig. \ref{Fig.4}(b). The applied magnetic field brings considerable polarization among the localized spins, and thereby the Mn spin system starts gaining the net magnetic moment, which is indicated by a small arrow (purple) in the upper panel of fig. \ref{Fig.4}(b). The hole either occupies a spin-up or spin-down level according to the initial Zeeman splitting of the energy levels. The lower panel indicates the preferential direction of the hole spin to be along the direction of the B, due to which the net magnetic field $\mathrm{B_{net}=B+B_{exc}}$ acting on the spin system grows, which results in an increase of the z component of the Mn spins indicated by the large size arrow in fig. \ref{Fig.4}(b). Our results report the larger polaron binding energy due to the confinement in QR compared to the values obtained with other quantum systems \cite{mackh1993exciton, wojnar2008size}. The reliability of the results can be confirmed only via experimental verifications to understand the internal mechanism that occurs in the QR responsible for such a larger EMP binding energy.\\
\indent Turning now to configuration (ii), fig. \ref{Fig.5}(a) plots the variation of EMP energy as a function of magnetic field for the simultaneous increase of both the $\mathrm{x_{in}}$ and $\mathrm{x_{out}}$. The trend followed by $\mathrm{\Delta E}$ with B is as same as the trend discussed with configuration (i) above, but with a different suppression effectiveness and with a following exception: In contrast to configuration (i), the polaron energy increases with the increase of the $\mathrm{Mn^{2+}}$ dopants beyond the dilute limit both in the ring and the barrier regimes, because now the hole wavefunction can interact with a large number of Mn ions resides in the same layer and the barrier. Not only that, the exciton effective Bohr radius decreases with the increase of Mn contents which causes the squeezing of the hole wavefunction leading to a robust localization inside the ring that strongly favours the internal exchange interaction. This increase in polaron energy is also attributed to the input from the antiferromagnetic (AF) spin clusters, which increases with the addition of Mn contents. Experimental results \cite{mackh1994localized} show that the response of the AF spin clusters to a strongly non-uniform exchange molecular field induced by the localized carrier for the increased value of the polaron energy, though the input from the isolated neighboring para spins to the MP energy is reduced due to the AF interaction.\\ 
\indent As discussed above, the polaron suppression is governed by the degree of spin alignment in the global spin system of Mn ions with the external magnetic field. The slope of the circular polarization degree in low magnetic field varies, as shown in fig. \ref{Fig.5}(c) for various Mn contents inside the ring maintained at T= 4.2K. While the external field strength of B = 1.2T is able to suppress $\mathrm{\Delta E}$ for $\mathrm{x_{in}=0.005}$ entirely, the former is shifted to higher values for larger Mn contents because of the magnetic hardening of the Mn spin system though a slow decrease of polaron energy is observed within the field range of B = 1.2T. This is attributed to the increase of the internal exchange field, $\mathrm{B_{exc}}$ (fig. \ref{Fig.5}(d)) for larger values of x, and the respective decrease of the hole localization radius ($\mathrm{r_{MP}}$) from 29\AA \, to 11.5\AA \, (fig. \ref{Fig.5}(d)).
\section{Summary \label{summary}}
\indent Theoretical calculations have been presented for the magnetic polaron energies in QR with two different dopant configurations. As expected, the confinement provided by the QR both the radially and axially direction enhances the sp-d exchange interaction and, in turn, aids the formation of the EMP. The calculations of the magnetic field dependence of the magnetic polaron energy are in good qualitative agreement with the experimental observations that have been made with the bulk, 2D, and 0D quantum systems. Depending on the presence and absence of molar Mn contents in a particular layer of the ring topology, two contradictory trends of the polaron energy is observed. While configuration (i) reduces the polaronic energy shift at 0T by the AF interactions between the nearest para spins, the enhancement of polaronic energy shift is observed with configuration (ii) due to the presence of Mn ions both in the ring and barrier regimes. Such tuning of the exciton-bound states with the concentration of magnetic ions would be responsible for exciting spin phenomena that could be exploited in various spin-based devices. Time evolution dynamics of the MP have yet to be studied to understand the interplay between the exciton lifetime and the polaron formation time. But, in our earlier studies, we found a long exciton lifetime in QR when compared to other existing quantum systems, which predicts that the polaron formation time is very short so that it cannot be interrupted by the recombination process. The existence of toroidal topology by the QR makes carriers more responsive to magnetic perturbations, and therefore integrating the QR with the DMS would bring fascinating physics in nanomagnetism and spintronics. This work would stimulate the experimental studies of the MP formation in DMS QR for better understanding, and these unique properties of the system could be exploited for novel applications in optoelectronic and spintronic devices.

\bibliography{main}

\begin{thebibliography}{47}%
\makeatletter
\providecommand \@ifxundefined [1]{%
 \@ifx{#1\undefined}
}%
\providecommand \@ifnum [1]{%
 \ifnum #1\expandafter \@firstoftwo
 \else \expandafter \@secondoftwo
 \fi
}%
\providecommand \@ifx [1]{%
 \ifx #1\expandafter \@firstoftwo
 \else \expandafter \@secondoftwo
 \fi
}%
\providecommand \natexlab [1]{#1}%
\providecommand \enquote  [1]{``#1''}%
\providecommand \bibnamefont  [1]{#1}%
\providecommand \bibfnamefont [1]{#1}%
\providecommand \citenamefont [1]{#1}%
\providecommand \href@noop [0]{\@secondoftwo}%
\providecommand \href [0]{\begingroup \@sanitize@url \@href}%
\providecommand \@href[1]{\@@startlink{#1}\@@href}%
\providecommand \@@href[1]{\endgroup#1\@@endlink}%
\providecommand \@sanitize@url [0]{\catcode `\\12\catcode `\$12\catcode
  `\&12\catcode `\#12\catcode `\^12\catcode `\_12\catcode `\%12\relax}%
\providecommand \@@startlink[1]{}%
\providecommand \@@endlink[0]{}%
\providecommand \url  [0]{\begingroup\@sanitize@url \@url }%
\providecommand \@url [1]{\endgroup\@href {#1}{\urlprefix }}%
\providecommand \urlprefix  [0]{URL }%
\providecommand \Eprint [0]{\href }%
\providecommand \doibase [0]{https://doi.org/}%
\providecommand \selectlanguage [0]{\@gobble}%
\providecommand \bibinfo  [0]{\@secondoftwo}%
\providecommand \bibfield  [0]{\@secondoftwo}%
\providecommand \translation [1]{[#1]}%
\providecommand \BibitemOpen [0]{}%
\providecommand \bibitemStop [0]{}%
\providecommand \bibitemNoStop [0]{.\EOS\space}%
\providecommand \EOS [0]{\spacefactor3000\relax}%
\providecommand \BibitemShut  [1]{\csname bibitem#1\endcsname}%
\let\auto@bib@innerbib\@empty
\bibitem [{\citenamefont {Furdyna}(1988)}]{furdyna1988diluted}%
  \BibitemOpen
  \bibfield  {author} {\bibinfo {author} {\bibfnamefont {J.~K.}\ \bibnamefont
  {Furdyna}},\ }\href@noop {} {\bibfield  {journal} {\bibinfo  {journal}
  {Journal of Applied Physics}\ }\textbf {\bibinfo {volume} {64}},\ \bibinfo
  {pages} {R29} (\bibinfo {year} {1988})}\BibitemShut {NoStop}%
\bibitem [{\citenamefont {Gaj}\ \emph {et~al.}(1993{\natexlab{a}})\citenamefont
  {Gaj}, \citenamefont {Planel},\ and\ \citenamefont
  {Fishman}}]{gaj1993relation}%
  \BibitemOpen
  \bibfield  {author} {\bibinfo {author} {\bibfnamefont {J.}~\bibnamefont
  {Gaj}}, \bibinfo {author} {\bibfnamefont {R.}~\bibnamefont {Planel}},\ and\
  \bibinfo {author} {\bibfnamefont {G.}~\bibnamefont {Fishman}},\ }\href@noop
  {} {\bibfield  {journal} {\bibinfo  {journal} {Solid State Communications}\
  }\textbf {\bibinfo {volume} {88}},\ \bibinfo {pages} {927} (\bibinfo {year}
  {1993}{\natexlab{a}})}\BibitemShut {NoStop}%
\bibitem [{\citenamefont {Kalpana}\ \emph {et~al.}(2017)\citenamefont
  {Kalpana}, \citenamefont {Nithiananthi},\ and\ \citenamefont
  {Jayakumar}}]{kalpana2017donor}%
  \BibitemOpen
  \bibfield  {author} {\bibinfo {author} {\bibfnamefont {P.~S.}\ \bibnamefont
  {Kalpana}}, \bibinfo {author} {\bibfnamefont {P.}~\bibnamefont
  {Nithiananthi}},\ and\ \bibinfo {author} {\bibfnamefont {K.}~\bibnamefont
  {Jayakumar}},\ }\href@noop {} {\bibfield  {journal} {\bibinfo  {journal}
  {Superlattices and Microstructures}\ }\textbf {\bibinfo {volume} {102}},\
  \bibinfo {pages} {246} (\bibinfo {year} {2017})}\BibitemShut {NoStop}%
\bibitem [{\citenamefont {Rice}\ \emph {et~al.}(2017)\citenamefont {Rice},
  \citenamefont {Liu}, \citenamefont {Pinchetti}, \citenamefont {Yakovlev},
  \citenamefont {Klimov},\ and\ \citenamefont {Crooker}}]{rice2017direct}%
  \BibitemOpen
  \bibfield  {author} {\bibinfo {author} {\bibfnamefont {W.}~\bibnamefont
  {Rice}}, \bibinfo {author} {\bibfnamefont {W.}~\bibnamefont {Liu}}, \bibinfo
  {author} {\bibfnamefont {V.}~\bibnamefont {Pinchetti}}, \bibinfo {author}
  {\bibfnamefont {D.}~\bibnamefont {Yakovlev}}, \bibinfo {author}
  {\bibfnamefont {V.}~\bibnamefont {Klimov}},\ and\ \bibinfo {author}
  {\bibfnamefont {S.}~\bibnamefont {Crooker}},\ }\href@noop {} {\bibfield
  {journal} {\bibinfo  {journal} {Nano letters}\ }\textbf {\bibinfo {volume}
  {17}},\ \bibinfo {pages} {3068} (\bibinfo {year} {2017})}\BibitemShut
  {NoStop}%
\bibitem [{\citenamefont {Fainblat}\ \emph {et~al.}(2016)\citenamefont
  {Fainblat}, \citenamefont {Barrows}, \citenamefont {Hopmann}, \citenamefont
  {Siebeneicher}, \citenamefont {Vlaskin}, \citenamefont {Gamelin},\ and\
  \citenamefont {Bacher}}]{fainblat2016giant}%
  \BibitemOpen
  \bibfield  {author} {\bibinfo {author} {\bibfnamefont {R.}~\bibnamefont
  {Fainblat}}, \bibinfo {author} {\bibfnamefont {C.~J.}\ \bibnamefont
  {Barrows}}, \bibinfo {author} {\bibfnamefont {E.}~\bibnamefont {Hopmann}},
  \bibinfo {author} {\bibfnamefont {S.}~\bibnamefont {Siebeneicher}}, \bibinfo
  {author} {\bibfnamefont {V.~A.}\ \bibnamefont {Vlaskin}}, \bibinfo {author}
  {\bibfnamefont {D.~R.}\ \bibnamefont {Gamelin}},\ and\ \bibinfo {author}
  {\bibfnamefont {G.}~\bibnamefont {Bacher}},\ }\href@noop {} {\bibfield
  {journal} {\bibinfo  {journal} {Nano letters}\ }\textbf {\bibinfo {volume}
  {16}},\ \bibinfo {pages} {6371} (\bibinfo {year} {2016})}\BibitemShut
  {NoStop}%
\bibitem [{\citenamefont {Barrows}\ \emph {et~al.}(2017)\citenamefont
  {Barrows}, \citenamefont {Fainblat},\ and\ \citenamefont
  {Gamelin}}]{barrows2017excitonic}%
  \BibitemOpen
  \bibfield  {author} {\bibinfo {author} {\bibfnamefont {C.~J.}\ \bibnamefont
  {Barrows}}, \bibinfo {author} {\bibfnamefont {R.}~\bibnamefont {Fainblat}},\
  and\ \bibinfo {author} {\bibfnamefont {D.~R.}\ \bibnamefont {Gamelin}},\
  }\href@noop {} {\bibfield  {journal} {\bibinfo  {journal} {Journal of
  Materials Chemistry C}\ }\textbf {\bibinfo {volume} {5}},\ \bibinfo {pages}
  {5232} (\bibinfo {year} {2017})}\BibitemShut {NoStop}%
\bibitem [{\citenamefont {Gaj}\ \emph {et~al.}(1993{\natexlab{b}})\citenamefont
  {Gaj}, \citenamefont {Gatazka},\ and\ \citenamefont
  {Nawrocki}}]{gaj1993giant}%
  \BibitemOpen
  \bibfield  {author} {\bibinfo {author} {\bibfnamefont {J.}~\bibnamefont
  {Gaj}}, \bibinfo {author} {\bibfnamefont {R.}~\bibnamefont {Gatazka}},\ and\
  \bibinfo {author} {\bibfnamefont {M.}~\bibnamefont {Nawrocki}},\ }\href@noop
  {} {\bibfield  {journal} {\bibinfo  {journal} {Solid State Communications}\
  }\textbf {\bibinfo {volume} {88}},\ \bibinfo {pages} {923} (\bibinfo {year}
  {1993}{\natexlab{b}})}\BibitemShut {NoStop}%
\bibitem [{\citenamefont {Panmand}\ \emph {et~al.}(2020)\citenamefont
  {Panmand}, \citenamefont {Tekale}, \citenamefont {Daware}, \citenamefont
  {Gosavi}, \citenamefont {Jha},\ and\ \citenamefont
  {Kale}}]{panmand2020characterisation}%
  \BibitemOpen
  \bibfield  {author} {\bibinfo {author} {\bibfnamefont {R.~P.}\ \bibnamefont
  {Panmand}}, \bibinfo {author} {\bibfnamefont {S.~P.}\ \bibnamefont {Tekale}},
  \bibinfo {author} {\bibfnamefont {K.~D.}\ \bibnamefont {Daware}}, \bibinfo
  {author} {\bibfnamefont {S.~W.}\ \bibnamefont {Gosavi}}, \bibinfo {author}
  {\bibfnamefont {A.}~\bibnamefont {Jha}},\ and\ \bibinfo {author}
  {\bibfnamefont {B.~B.}\ \bibnamefont {Kale}},\ }\href@noop {} {\bibfield
  {journal} {\bibinfo  {journal} {Journal of Alloys and Compounds}\ }\textbf
  {\bibinfo {volume} {817}},\ \bibinfo {pages} {152696} (\bibinfo {year}
  {2020})}\BibitemShut {NoStop}%
\bibitem [{\citenamefont {Rossin}\ \emph {et~al.}(1996)\citenamefont {Rossin},
  \citenamefont {Henneberger},\ and\ \citenamefont
  {Puls}}]{rossin1996magnetic}%
  \BibitemOpen
  \bibfield  {author} {\bibinfo {author} {\bibfnamefont {V.}~\bibnamefont
  {Rossin}}, \bibinfo {author} {\bibfnamefont {F.}~\bibnamefont
  {Henneberger}},\ and\ \bibinfo {author} {\bibfnamefont {J.}~\bibnamefont
  {Puls}},\ }\href@noop {} {\bibfield  {journal} {\bibinfo  {journal} {Physical
  Review B}\ }\textbf {\bibinfo {volume} {53}},\ \bibinfo {pages} {16444}
  (\bibinfo {year} {1996})}\BibitemShut {NoStop}%
\bibitem [{\citenamefont {Akimov}\ \emph {et~al.}(2017)\citenamefont {Akimov},
  \citenamefont {Godde}, \citenamefont {Kavokin}, \citenamefont {Yakovlev},
  \citenamefont {Reshina}, \citenamefont {Sedova}, \citenamefont {Sorokin},
  \citenamefont {Ivanov}, \citenamefont {Kusrayev},\ and\ \citenamefont
  {Bayer}}]{akimov2017dynamics}%
  \BibitemOpen
  \bibfield  {author} {\bibinfo {author} {\bibfnamefont {I.}~\bibnamefont
  {Akimov}}, \bibinfo {author} {\bibfnamefont {T.}~\bibnamefont {Godde}},
  \bibinfo {author} {\bibfnamefont {K.}~\bibnamefont {Kavokin}}, \bibinfo
  {author} {\bibfnamefont {D.}~\bibnamefont {Yakovlev}}, \bibinfo {author}
  {\bibfnamefont {I.}~\bibnamefont {Reshina}}, \bibinfo {author} {\bibfnamefont
  {I.}~\bibnamefont {Sedova}}, \bibinfo {author} {\bibfnamefont
  {S.}~\bibnamefont {Sorokin}}, \bibinfo {author} {\bibfnamefont
  {S.}~\bibnamefont {Ivanov}}, \bibinfo {author} {\bibfnamefont {Y.~G.}\
  \bibnamefont {Kusrayev}},\ and\ \bibinfo {author} {\bibfnamefont
  {M.}~\bibnamefont {Bayer}},\ }\href@noop {} {\bibfield  {journal} {\bibinfo
  {journal} {Physical Review B}\ }\textbf {\bibinfo {volume} {95}},\ \bibinfo
  {pages} {155303} (\bibinfo {year} {2017})}\BibitemShut {NoStop}%
\bibitem [{\citenamefont {Kalpana}\ and\ \citenamefont
  {Jayakumar}(2017)}]{kalpana2017bound}%
  \BibitemOpen
  \bibfield  {author} {\bibinfo {author} {\bibfnamefont {P.}~\bibnamefont
  {Kalpana}}\ and\ \bibinfo {author} {\bibfnamefont {K.}~\bibnamefont
  {Jayakumar}},\ }\href@noop {} {\bibfield  {journal} {\bibinfo  {journal}
  {Physica E: Low-dimensional Systems and Nanostructures}\ }\textbf {\bibinfo
  {volume} {93}},\ \bibinfo {pages} {252} (\bibinfo {year} {2017})}\BibitemShut
  {NoStop}%
\bibitem [{\citenamefont {Kozyrev}\ \emph {et~al.}(2021)\citenamefont
  {Kozyrev}, \citenamefont {Akhmadullin}, \citenamefont {Namozov},
  \citenamefont {Kusrayev}, \citenamefont {Karczewski},\ and\ \citenamefont
  {Wojtowicz}}]{kozyrev2021optical}%
  \BibitemOpen
  \bibfield  {author} {\bibinfo {author} {\bibfnamefont {N.}~\bibnamefont
  {Kozyrev}}, \bibinfo {author} {\bibfnamefont {R.}~\bibnamefont
  {Akhmadullin}}, \bibinfo {author} {\bibfnamefont {B.}~\bibnamefont
  {Namozov}}, \bibinfo {author} {\bibfnamefont {Y.~G.}\ \bibnamefont
  {Kusrayev}}, \bibinfo {author} {\bibfnamefont {G.}~\bibnamefont
  {Karczewski}},\ and\ \bibinfo {author} {\bibfnamefont {T.}~\bibnamefont
  {Wojtowicz}},\ }\href@noop {} {\bibfield  {journal} {\bibinfo  {journal}
  {Physical Review B}\ }\textbf {\bibinfo {volume} {104}},\ \bibinfo {pages}
  {045307} (\bibinfo {year} {2021})}\BibitemShut {NoStop}%
\bibitem [{\citenamefont {Seufert}\ \emph {et~al.}(2001)\citenamefont
  {Seufert}, \citenamefont {Bacher}, \citenamefont {Scheibner}, \citenamefont
  {Forchel}, \citenamefont {Lee}, \citenamefont {Dobrowolska},\ and\
  \citenamefont {Furdyna}}]{seufert2001dynamical}%
  \BibitemOpen
  \bibfield  {author} {\bibinfo {author} {\bibfnamefont {J.}~\bibnamefont
  {Seufert}}, \bibinfo {author} {\bibfnamefont {G.}~\bibnamefont {Bacher}},
  \bibinfo {author} {\bibfnamefont {M.}~\bibnamefont {Scheibner}}, \bibinfo
  {author} {\bibfnamefont {A.}~\bibnamefont {Forchel}}, \bibinfo {author}
  {\bibfnamefont {S.}~\bibnamefont {Lee}}, \bibinfo {author} {\bibfnamefont
  {M.}~\bibnamefont {Dobrowolska}},\ and\ \bibinfo {author} {\bibfnamefont
  {J.}~\bibnamefont {Furdyna}},\ }\href@noop {} {\bibfield  {journal} {\bibinfo
   {journal} {Physical review letters}\ }\textbf {\bibinfo {volume} {88}},\
  \bibinfo {pages} {027402} (\bibinfo {year} {2001})}\BibitemShut {NoStop}%
\bibitem [{\citenamefont {Barman}\ \emph {et~al.}(2015)\citenamefont {Barman},
  \citenamefont {Oszwa{\l}dowski}, \citenamefont {Schweidenback}, \citenamefont
  {Russ}, \citenamefont {Pientka}, \citenamefont {Tsai}, \citenamefont {Chou},
  \citenamefont {Fan}, \citenamefont {Murphy}, \citenamefont {Cartwright} \emph
  {et~al.}}]{barman2015time}%
  \BibitemOpen
  \bibfield  {author} {\bibinfo {author} {\bibfnamefont {B.}~\bibnamefont
  {Barman}}, \bibinfo {author} {\bibfnamefont {R.}~\bibnamefont
  {Oszwa{\l}dowski}}, \bibinfo {author} {\bibfnamefont {L.}~\bibnamefont
  {Schweidenback}}, \bibinfo {author} {\bibfnamefont {A.}~\bibnamefont {Russ}},
  \bibinfo {author} {\bibfnamefont {J.}~\bibnamefont {Pientka}}, \bibinfo
  {author} {\bibfnamefont {Y.}~\bibnamefont {Tsai}}, \bibinfo {author}
  {\bibfnamefont {W.-C.}\ \bibnamefont {Chou}}, \bibinfo {author}
  {\bibfnamefont {W.}~\bibnamefont {Fan}}, \bibinfo {author} {\bibfnamefont
  {J.}~\bibnamefont {Murphy}}, \bibinfo {author} {\bibfnamefont
  {A.}~\bibnamefont {Cartwright}}, \emph {et~al.},\ }\href@noop {} {\bibfield
  {journal} {\bibinfo  {journal} {Physical Review B}\ }\textbf {\bibinfo
  {volume} {92}},\ \bibinfo {pages} {035430} (\bibinfo {year}
  {2015})}\BibitemShut {NoStop}%
\bibitem [{\citenamefont {Gurung}\ \emph {et~al.}(2008)\citenamefont {Gurung},
  \citenamefont {Mackowski}, \citenamefont {Karczewski}, \citenamefont
  {Jackson},\ and\ \citenamefont {Smith}}]{gurung2008ultralong}%
  \BibitemOpen
  \bibfield  {author} {\bibinfo {author} {\bibfnamefont {T.}~\bibnamefont
  {Gurung}}, \bibinfo {author} {\bibfnamefont {S.}~\bibnamefont {Mackowski}},
  \bibinfo {author} {\bibfnamefont {G.}~\bibnamefont {Karczewski}}, \bibinfo
  {author} {\bibfnamefont {H.~E.}\ \bibnamefont {Jackson}},\ and\ \bibinfo
  {author} {\bibfnamefont {L.~M.}\ \bibnamefont {Smith}},\ }\href@noop {}
  {\bibfield  {journal} {\bibinfo  {journal} {Applied Physics Letters}\
  }\textbf {\bibinfo {volume} {93}},\ \bibinfo {pages} {153114} (\bibinfo
  {year} {2008})}\BibitemShut {NoStop}%
\bibitem [{\citenamefont {Xiu}\ \emph {et~al.}(2010)\citenamefont {Xiu},
  \citenamefont {Wang}, \citenamefont {Kim}, \citenamefont {Upadhyaya},
  \citenamefont {Zhou}, \citenamefont {Kou}, \citenamefont {Han}, \citenamefont
  {Kawakami}, \citenamefont {Zou},\ and\ \citenamefont {Wang}}]{xiu2010room}%
  \BibitemOpen
  \bibfield  {author} {\bibinfo {author} {\bibfnamefont {F.}~\bibnamefont
  {Xiu}}, \bibinfo {author} {\bibfnamefont {Y.}~\bibnamefont {Wang}}, \bibinfo
  {author} {\bibfnamefont {J.}~\bibnamefont {Kim}}, \bibinfo {author}
  {\bibfnamefont {P.}~\bibnamefont {Upadhyaya}}, \bibinfo {author}
  {\bibfnamefont {Y.}~\bibnamefont {Zhou}}, \bibinfo {author} {\bibfnamefont
  {X.}~\bibnamefont {Kou}}, \bibinfo {author} {\bibfnamefont {W.}~\bibnamefont
  {Han}}, \bibinfo {author} {\bibfnamefont {R.}~\bibnamefont {Kawakami}},
  \bibinfo {author} {\bibfnamefont {J.}~\bibnamefont {Zou}},\ and\ \bibinfo
  {author} {\bibfnamefont {K.~L.}\ \bibnamefont {Wang}},\ }\href@noop {}
  {\bibfield  {journal} {\bibinfo  {journal} {ACS nano}\ }\textbf {\bibinfo
  {volume} {4}},\ \bibinfo {pages} {4948} (\bibinfo {year} {2010})}\BibitemShut
  {NoStop}%
\bibitem [{\citenamefont {Takeyama}\ \emph {et~al.}(1995)\citenamefont
  {Takeyama}, \citenamefont {Adachi}, \citenamefont {Takagi},\ and\
  \citenamefont {Aguekian}}]{takeyama1995exciton}%
  \BibitemOpen
  \bibfield  {author} {\bibinfo {author} {\bibfnamefont {S.}~\bibnamefont
  {Takeyama}}, \bibinfo {author} {\bibfnamefont {S.}~\bibnamefont {Adachi}},
  \bibinfo {author} {\bibfnamefont {Y.}~\bibnamefont {Takagi}},\ and\ \bibinfo
  {author} {\bibfnamefont {V.}~\bibnamefont {Aguekian}},\ }\href@noop {}
  {\bibfield  {journal} {\bibinfo  {journal} {Physical Review B}\ }\textbf
  {\bibinfo {volume} {51}},\ \bibinfo {pages} {4858} (\bibinfo {year}
  {1995})}\BibitemShut {NoStop}%
\bibitem [{\citenamefont {Zhukov}\ \emph {et~al.}(2019)\citenamefont {Zhukov},
  \citenamefont {Kusrayev}, \citenamefont {Kirstein}, \citenamefont {Thomann},
  \citenamefont {Salewski}, \citenamefont {Kozyrev}, \citenamefont {Yakovlev},\
  and\ \citenamefont {Bayer}}]{zhukov2019optical}%
  \BibitemOpen
  \bibfield  {author} {\bibinfo {author} {\bibfnamefont {E.}~\bibnamefont
  {Zhukov}}, \bibinfo {author} {\bibfnamefont {Y.~G.}\ \bibnamefont
  {Kusrayev}}, \bibinfo {author} {\bibfnamefont {E.}~\bibnamefont {Kirstein}},
  \bibinfo {author} {\bibfnamefont {A.}~\bibnamefont {Thomann}}, \bibinfo
  {author} {\bibfnamefont {M.}~\bibnamefont {Salewski}}, \bibinfo {author}
  {\bibfnamefont {N.}~\bibnamefont {Kozyrev}}, \bibinfo {author} {\bibfnamefont
  {D.}~\bibnamefont {Yakovlev}},\ and\ \bibinfo {author} {\bibfnamefont
  {M.}~\bibnamefont {Bayer}},\ }\href@noop {} {\bibfield  {journal} {\bibinfo
  {journal} {Physical Review B}\ }\textbf {\bibinfo {volume} {99}},\ \bibinfo
  {pages} {115204} (\bibinfo {year} {2019})}\BibitemShut {NoStop}%
\bibitem [{\citenamefont {Elangovan}\ \emph {et~al.}(1994)\citenamefont
  {Elangovan}, \citenamefont {Shyamada},\ and\ \citenamefont
  {Navaneethakrishnan}}]{elangovan1994bound}%
  \BibitemOpen
  \bibfield  {author} {\bibinfo {author} {\bibfnamefont {A.}~\bibnamefont
  {Elangovan}}, \bibinfo {author} {\bibfnamefont {D.}~\bibnamefont
  {Shyamada}},\ and\ \bibinfo {author} {\bibfnamefont {K.}~\bibnamefont
  {Navaneethakrishnan}},\ }\href@noop {} {\bibfield  {journal} {\bibinfo
  {journal} {Solid state communications}\ }\textbf {\bibinfo {volume} {89}},\
  \bibinfo {pages} {869} (\bibinfo {year} {1994})}\BibitemShut {NoStop}%
\bibitem [{\citenamefont {Mackh}\ \emph
  {et~al.}(1994{\natexlab{a}})\citenamefont {Mackh}, \citenamefont {Ossau},
  \citenamefont {Yakovlev}, \citenamefont {Waag}, \citenamefont {Landwehr},
  \citenamefont {Hellmann},\ and\ \citenamefont
  {G{\"o}bel}}]{mackh1994localized}%
  \BibitemOpen
  \bibfield  {author} {\bibinfo {author} {\bibfnamefont {G.}~\bibnamefont
  {Mackh}}, \bibinfo {author} {\bibfnamefont {W.}~\bibnamefont {Ossau}},
  \bibinfo {author} {\bibfnamefont {D.}~\bibnamefont {Yakovlev}}, \bibinfo
  {author} {\bibfnamefont {A.}~\bibnamefont {Waag}}, \bibinfo {author}
  {\bibfnamefont {G.}~\bibnamefont {Landwehr}}, \bibinfo {author}
  {\bibfnamefont {R.}~\bibnamefont {Hellmann}},\ and\ \bibinfo {author}
  {\bibfnamefont {E.}~\bibnamefont {G{\"o}bel}},\ }\href@noop {} {\bibfield
  {journal} {\bibinfo  {journal} {Physical Review B}\ }\textbf {\bibinfo
  {volume} {49}},\ \bibinfo {pages} {10248} (\bibinfo {year}
  {1994}{\natexlab{a}})}\BibitemShut {NoStop}%
\bibitem [{\citenamefont {Yakovlev}\ \emph {et~al.}(1990)\citenamefont
  {Yakovlev}, \citenamefont {Ossau}, \citenamefont {Landwehr}, \citenamefont
  {Bicknell-Tassius}, \citenamefont {Waag},\ and\ \citenamefont
  {Uraltsev}}]{yakovlev1990first}%
  \BibitemOpen
  \bibfield  {author} {\bibinfo {author} {\bibfnamefont {D.}~\bibnamefont
  {Yakovlev}}, \bibinfo {author} {\bibfnamefont {W.}~\bibnamefont {Ossau}},
  \bibinfo {author} {\bibfnamefont {G.}~\bibnamefont {Landwehr}}, \bibinfo
  {author} {\bibfnamefont {R.}~\bibnamefont {Bicknell-Tassius}}, \bibinfo
  {author} {\bibfnamefont {A.}~\bibnamefont {Waag}},\ and\ \bibinfo {author}
  {\bibfnamefont {I.}~\bibnamefont {Uraltsev}},\ }\href@noop {} {\bibfield
  {journal} {\bibinfo  {journal} {Solid state communications}\ }\textbf
  {\bibinfo {volume} {76}},\ \bibinfo {pages} {325} (\bibinfo {year}
  {1990})}\BibitemShut {NoStop}%
\bibitem [{\citenamefont {Benoit {\`A} La~Guillaume}(1993)}]{benoit1993free}%
  \BibitemOpen
  \bibfield  {author} {\bibinfo {author} {\bibfnamefont {C.}~\bibnamefont
  {Benoit {\`A} La~Guillaume}},\ }\href@noop {} {\bibfield  {journal} {\bibinfo
   {journal} {physica status solidi (b)}\ }\textbf {\bibinfo {volume} {175}},\
  \bibinfo {pages} {369} (\bibinfo {year} {1993})}\BibitemShut {NoStop}%
\bibitem [{\citenamefont {Mackh}\ \emph {et~al.}(1993)\citenamefont {Mackh},
  \citenamefont {Ossau}, \citenamefont {Yakovlev}, \citenamefont {Waag},
  \citenamefont {Litz},\ and\ \citenamefont {Landwehr}}]{mackh1993exciton}%
  \BibitemOpen
  \bibfield  {author} {\bibinfo {author} {\bibfnamefont {G.}~\bibnamefont
  {Mackh}}, \bibinfo {author} {\bibfnamefont {W.}~\bibnamefont {Ossau}},
  \bibinfo {author} {\bibfnamefont {D.}~\bibnamefont {Yakovlev}}, \bibinfo
  {author} {\bibfnamefont {A.}~\bibnamefont {Waag}}, \bibinfo {author}
  {\bibfnamefont {T.}~\bibnamefont {Litz}},\ and\ \bibinfo {author}
  {\bibfnamefont {G.}~\bibnamefont {Landwehr}},\ }\href@noop {} {\bibfield
  {journal} {\bibinfo  {journal} {Solid state communications}\ }\textbf
  {\bibinfo {volume} {88}},\ \bibinfo {pages} {221} (\bibinfo {year}
  {1993})}\BibitemShut {NoStop}%
\bibitem [{\citenamefont {Poweleit}\ \emph {et~al.}(1994)\citenamefont
  {Poweleit}, \citenamefont {Smith},\ and\ \citenamefont
  {Jonker}}]{poweleit1994observation}%
  \BibitemOpen
  \bibfield  {author} {\bibinfo {author} {\bibfnamefont {C.}~\bibnamefont
  {Poweleit}}, \bibinfo {author} {\bibfnamefont {L.}~\bibnamefont {Smith}},\
  and\ \bibinfo {author} {\bibfnamefont {B.}~\bibnamefont {Jonker}},\
  }\href@noop {} {\bibfield  {journal} {\bibinfo  {journal} {Physical Review
  B}\ }\textbf {\bibinfo {volume} {50}},\ \bibinfo {pages} {18662} (\bibinfo
  {year} {1994})}\BibitemShut {NoStop}%
\bibitem [{\citenamefont {Mackh}\ \emph
  {et~al.}(1994{\natexlab{b}})\citenamefont {Mackh}, \citenamefont {Hilpert},
  \citenamefont {Yakovlev}, \citenamefont {Ossau}, \citenamefont {Heinke},
  \citenamefont {Litz}, \citenamefont {Fischer}, \citenamefont {Waag},
  \citenamefont {Landwehr}, \citenamefont {Hellmann} \emph
  {et~al.}}]{mackh1994exciton}%
  \BibitemOpen
  \bibfield  {author} {\bibinfo {author} {\bibfnamefont {G.}~\bibnamefont
  {Mackh}}, \bibinfo {author} {\bibfnamefont {M.}~\bibnamefont {Hilpert}},
  \bibinfo {author} {\bibfnamefont {D.}~\bibnamefont {Yakovlev}}, \bibinfo
  {author} {\bibfnamefont {W.}~\bibnamefont {Ossau}}, \bibinfo {author}
  {\bibfnamefont {H.}~\bibnamefont {Heinke}}, \bibinfo {author} {\bibfnamefont
  {T.}~\bibnamefont {Litz}}, \bibinfo {author} {\bibfnamefont {F.}~\bibnamefont
  {Fischer}}, \bibinfo {author} {\bibfnamefont {A.}~\bibnamefont {Waag}},
  \bibinfo {author} {\bibfnamefont {G.}~\bibnamefont {Landwehr}}, \bibinfo
  {author} {\bibfnamefont {R.}~\bibnamefont {Hellmann}}, \emph {et~al.},\
  }\href@noop {} {\bibfield  {journal} {\bibinfo  {journal} {Physical Review
  B}\ }\textbf {\bibinfo {volume} {50}},\ \bibinfo {pages} {14069} (\bibinfo
  {year} {1994}{\natexlab{b}})}\BibitemShut {NoStop}%
\bibitem [{\citenamefont {Zayhowski}\ \emph {et~al.}(1985)\citenamefont
  {Zayhowski}, \citenamefont {Jagannath}, \citenamefont {Kershaw},
  \citenamefont {Ridgley}, \citenamefont {Dwight},\ and\ \citenamefont
  {Wold}}]{zayhowski1985picosecond}%
  \BibitemOpen
  \bibfield  {author} {\bibinfo {author} {\bibfnamefont {J.}~\bibnamefont
  {Zayhowski}}, \bibinfo {author} {\bibfnamefont {C.}~\bibnamefont
  {Jagannath}}, \bibinfo {author} {\bibfnamefont {R.}~\bibnamefont {Kershaw}},
  \bibinfo {author} {\bibfnamefont {D.}~\bibnamefont {Ridgley}}, \bibinfo
  {author} {\bibfnamefont {K.}~\bibnamefont {Dwight}},\ and\ \bibinfo {author}
  {\bibfnamefont {A.}~\bibnamefont {Wold}},\ }\href@noop {} {\bibfield
  {journal} {\bibinfo  {journal} {Solid state communications}\ }\textbf
  {\bibinfo {volume} {55}},\ \bibinfo {pages} {941} (\bibinfo {year}
  {1985})}\BibitemShut {NoStop}%
\bibitem [{\citenamefont {Yakovlev}(1993)}]{yakovlev1993exciton}%
  \BibitemOpen
  \bibfield  {author} {\bibinfo {author} {\bibfnamefont {D.}~\bibnamefont
  {Yakovlev}},\ }\href@noop {} {\bibfield  {journal} {\bibinfo  {journal} {Le
  Journal de Physique IV}\ }\textbf {\bibinfo {volume} {3}},\ \bibinfo {pages}
  {C5} (\bibinfo {year} {1993})}\BibitemShut {NoStop}%
\bibitem [{\citenamefont {Yakovlev}\ \emph {et~al.}(1995)\citenamefont
  {Yakovlev}, \citenamefont {Mackh}, \citenamefont {Kuhn-Heinrich},
  \citenamefont {Ossau}, \citenamefont {Waag}, \citenamefont {Landwehr},
  \citenamefont {Hellmann},\ and\ \citenamefont
  {G{\"o}bel}}]{yakovlev1995exciton}%
  \BibitemOpen
  \bibfield  {author} {\bibinfo {author} {\bibfnamefont {D.}~\bibnamefont
  {Yakovlev}}, \bibinfo {author} {\bibfnamefont {G.}~\bibnamefont {Mackh}},
  \bibinfo {author} {\bibfnamefont {B.}~\bibnamefont {Kuhn-Heinrich}}, \bibinfo
  {author} {\bibfnamefont {W.}~\bibnamefont {Ossau}}, \bibinfo {author}
  {\bibfnamefont {A.}~\bibnamefont {Waag}}, \bibinfo {author} {\bibfnamefont
  {G.}~\bibnamefont {Landwehr}}, \bibinfo {author} {\bibfnamefont
  {R.}~\bibnamefont {Hellmann}},\ and\ \bibinfo {author} {\bibfnamefont
  {E.}~\bibnamefont {G{\"o}bel}},\ }\href@noop {} {\bibfield  {journal}
  {\bibinfo  {journal} {Physical Review B}\ }\textbf {\bibinfo {volume} {52}},\
  \bibinfo {pages} {12033} (\bibinfo {year} {1995})}\BibitemShut {NoStop}%
\bibitem [{\citenamefont {Zhukov}\ \emph {et~al.}(2016)\citenamefont {Zhukov},
  \citenamefont {Kusrayev}, \citenamefont {Kavokin}, \citenamefont {Yakovlev},
  \citenamefont {Debus}, \citenamefont {Schwan}, \citenamefont {Akimov},
  \citenamefont {Karczewski}, \citenamefont {Wojtowicz}, \citenamefont {Kossut}
  \emph {et~al.}}]{zhukov2016optical}%
  \BibitemOpen
  \bibfield  {author} {\bibinfo {author} {\bibfnamefont {E.}~\bibnamefont
  {Zhukov}}, \bibinfo {author} {\bibfnamefont {Y.~G.}\ \bibnamefont
  {Kusrayev}}, \bibinfo {author} {\bibfnamefont {K.}~\bibnamefont {Kavokin}},
  \bibinfo {author} {\bibfnamefont {D.}~\bibnamefont {Yakovlev}}, \bibinfo
  {author} {\bibfnamefont {J.}~\bibnamefont {Debus}}, \bibinfo {author}
  {\bibfnamefont {A.}~\bibnamefont {Schwan}}, \bibinfo {author} {\bibfnamefont
  {I.}~\bibnamefont {Akimov}}, \bibinfo {author} {\bibfnamefont
  {G.}~\bibnamefont {Karczewski}}, \bibinfo {author} {\bibfnamefont
  {T.}~\bibnamefont {Wojtowicz}}, \bibinfo {author} {\bibfnamefont
  {J.}~\bibnamefont {Kossut}}, \emph {et~al.},\ }\href@noop {} {\bibfield
  {journal} {\bibinfo  {journal} {Physical Review B}\ }\textbf {\bibinfo
  {volume} {93}},\ \bibinfo {pages} {245305} (\bibinfo {year}
  {2016})}\BibitemShut {NoStop}%
\bibitem [{\citenamefont {Maksimov}\ \emph {et~al.}(2000)\citenamefont
  {Maksimov}, \citenamefont {Bacher}, \citenamefont {McDonald}, \citenamefont
  {Kulakovskii}, \citenamefont {Forchel}, \citenamefont {Becker}, \citenamefont
  {Landwehr},\ and\ \citenamefont {Molenkamp}}]{maksimov2000magnetic}%
  \BibitemOpen
  \bibfield  {author} {\bibinfo {author} {\bibfnamefont {A.}~\bibnamefont
  {Maksimov}}, \bibinfo {author} {\bibfnamefont {G.}~\bibnamefont {Bacher}},
  \bibinfo {author} {\bibfnamefont {A.}~\bibnamefont {McDonald}}, \bibinfo
  {author} {\bibfnamefont {V.}~\bibnamefont {Kulakovskii}}, \bibinfo {author}
  {\bibfnamefont {A.}~\bibnamefont {Forchel}}, \bibinfo {author} {\bibfnamefont
  {C.}~\bibnamefont {Becker}}, \bibinfo {author} {\bibfnamefont
  {G.}~\bibnamefont {Landwehr}},\ and\ \bibinfo {author} {\bibfnamefont
  {L.}~\bibnamefont {Molenkamp}},\ }\href@noop {} {\bibfield  {journal}
  {\bibinfo  {journal} {Physical Review B}\ }\textbf {\bibinfo {volume} {62}},\
  \bibinfo {pages} {R7767} (\bibinfo {year} {2000})}\BibitemShut {NoStop}%
\bibitem [{\citenamefont {Gnanasekar}\ and\ \citenamefont
  {Navaneethakrishnan}(2004)}]{gnanasekar2004spin}%
  \BibitemOpen
  \bibfield  {author} {\bibinfo {author} {\bibfnamefont {K.}~\bibnamefont
  {Gnanasekar}}\ and\ \bibinfo {author} {\bibfnamefont {K.}~\bibnamefont
  {Navaneethakrishnan}},\ }\href@noop {} {\bibfield  {journal} {\bibinfo
  {journal} {Modern Physics Letters B}\ }\textbf {\bibinfo {volume} {18}},\
  \bibinfo {pages} {419} (\bibinfo {year} {2004})}\BibitemShut {NoStop}%
\bibitem [{\citenamefont {Wojnar}\ \emph {et~al.}(2008)\citenamefont {Wojnar},
  \citenamefont {Suffczy{\'n}ski}, \citenamefont {Kowalik}, \citenamefont
  {Golnik}, \citenamefont {Aleszkiewicz}, \citenamefont {Karczewski},\ and\
  \citenamefont {Kossut}}]{wojnar2008size}%
  \BibitemOpen
  \bibfield  {author} {\bibinfo {author} {\bibfnamefont {P.}~\bibnamefont
  {Wojnar}}, \bibinfo {author} {\bibfnamefont {J.}~\bibnamefont
  {Suffczy{\'n}ski}}, \bibinfo {author} {\bibfnamefont {K.}~\bibnamefont
  {Kowalik}}, \bibinfo {author} {\bibfnamefont {A.}~\bibnamefont {Golnik}},
  \bibinfo {author} {\bibfnamefont {M.}~\bibnamefont {Aleszkiewicz}}, \bibinfo
  {author} {\bibfnamefont {G.}~\bibnamefont {Karczewski}},\ and\ \bibinfo
  {author} {\bibfnamefont {J.}~\bibnamefont {Kossut}},\ }\href@noop {}
  {\bibfield  {journal} {\bibinfo  {journal} {Nanotechnology}\ }\textbf
  {\bibinfo {volume} {19}},\ \bibinfo {pages} {235403} (\bibinfo {year}
  {2008})}\BibitemShut {NoStop}%
\bibitem [{\citenamefont {Sellers}\ \emph {et~al.}(2010)\citenamefont
  {Sellers}, \citenamefont {Oszwa{\l}dowski}, \citenamefont {Whiteside},
  \citenamefont {Eginligil}, \citenamefont {Petrou}, \citenamefont {Zutic},
  \citenamefont {Chou}, \citenamefont {Fan}, \citenamefont {Petukhov},
  \citenamefont {Kim} \emph {et~al.}}]{sellers2010robust}%
  \BibitemOpen
  \bibfield  {author} {\bibinfo {author} {\bibfnamefont {I.}~\bibnamefont
  {Sellers}}, \bibinfo {author} {\bibfnamefont {R.}~\bibnamefont
  {Oszwa{\l}dowski}}, \bibinfo {author} {\bibfnamefont {V.}~\bibnamefont
  {Whiteside}}, \bibinfo {author} {\bibfnamefont {M.}~\bibnamefont
  {Eginligil}}, \bibinfo {author} {\bibfnamefont {A.}~\bibnamefont {Petrou}},
  \bibinfo {author} {\bibfnamefont {I.}~\bibnamefont {Zutic}}, \bibinfo
  {author} {\bibfnamefont {W.-C.}\ \bibnamefont {Chou}}, \bibinfo {author}
  {\bibfnamefont {W.}~\bibnamefont {Fan}}, \bibinfo {author} {\bibfnamefont
  {A.}~\bibnamefont {Petukhov}}, \bibinfo {author} {\bibfnamefont
  {S.}~\bibnamefont {Kim}}, \emph {et~al.},\ }\href@noop {} {\bibfield
  {journal} {\bibinfo  {journal} {Physical Review B}\ }\textbf {\bibinfo
  {volume} {82}},\ \bibinfo {pages} {195320} (\bibinfo {year}
  {2010})}\BibitemShut {NoStop}%
\bibitem [{\citenamefont {Anitha}\ and\ \citenamefont
  {Nithiananthi}(2019)}]{anitha2019dynamics}%
  \BibitemOpen
  \bibfield  {author} {\bibinfo {author} {\bibfnamefont {B.}~\bibnamefont
  {Anitha}}\ and\ \bibinfo {author} {\bibfnamefont {P.}~\bibnamefont
  {Nithiananthi}},\ }in\ \href@noop {} {\emph {\bibinfo {booktitle} {AIP
  Conference Proceedings}}},\ Vol.\ \bibinfo {volume} {2115}\ (\bibinfo
  {organization} {AIP Publishing LLC},\ \bibinfo {year} {2019})\ p.\ \bibinfo
  {pages} {030455}\BibitemShut {NoStop}%
\bibitem [{\citenamefont {Muckel}\ \emph {et~al.}(2017)\citenamefont {Muckel},
  \citenamefont {Barrows}, \citenamefont {Graf}, \citenamefont {Schmitz},
  \citenamefont {Erickson}, \citenamefont {Gamelin},\ and\ \citenamefont
  {Bacher}}]{muckel2017current}%
  \BibitemOpen
  \bibfield  {author} {\bibinfo {author} {\bibfnamefont {F.}~\bibnamefont
  {Muckel}}, \bibinfo {author} {\bibfnamefont {C.~J.}\ \bibnamefont {Barrows}},
  \bibinfo {author} {\bibfnamefont {A.}~\bibnamefont {Graf}}, \bibinfo {author}
  {\bibfnamefont {A.}~\bibnamefont {Schmitz}}, \bibinfo {author} {\bibfnamefont
  {C.~S.}\ \bibnamefont {Erickson}}, \bibinfo {author} {\bibfnamefont {D.~R.}\
  \bibnamefont {Gamelin}},\ and\ \bibinfo {author} {\bibfnamefont
  {G.}~\bibnamefont {Bacher}},\ }\href@noop {} {\bibfield  {journal} {\bibinfo
  {journal} {Nano letters}\ }\textbf {\bibinfo {volume} {17}},\ \bibinfo
  {pages} {4768} (\bibinfo {year} {2017})}\BibitemShut {NoStop}%
\bibitem [{\citenamefont {Barman}\ \emph {et~al.}(2020)\citenamefont {Barman},
  \citenamefont {Pientka}, \citenamefont {Murphy}, \citenamefont {Cartwright},
  \citenamefont {Chou}, \citenamefont {Fan}, \citenamefont {Oszwa{\l}dowski},\
  and\ \citenamefont {Petrou}}]{barman2020circular}%
  \BibitemOpen
  \bibfield  {author} {\bibinfo {author} {\bibfnamefont {B.}~\bibnamefont
  {Barman}}, \bibinfo {author} {\bibfnamefont {J.~M.}\ \bibnamefont {Pientka}},
  \bibinfo {author} {\bibfnamefont {J.~R.}\ \bibnamefont {Murphy}}, \bibinfo
  {author} {\bibfnamefont {A.~N.}\ \bibnamefont {Cartwright}}, \bibinfo
  {author} {\bibfnamefont {W.-C.}\ \bibnamefont {Chou}}, \bibinfo {author}
  {\bibfnamefont {W.-C.}\ \bibnamefont {Fan}}, \bibinfo {author} {\bibfnamefont
  {R.}~\bibnamefont {Oszwa{\l}dowski}},\ and\ \bibinfo {author} {\bibfnamefont
  {A.}~\bibnamefont {Petrou}},\ }\href@noop {} {\bibfield  {journal} {\bibinfo
  {journal} {The Journal of Physical Chemistry C}\ }\textbf {\bibinfo {volume}
  {124}},\ \bibinfo {pages} {12766} (\bibinfo {year} {2020})}\BibitemShut
  {NoStop}%
\bibitem [{\citenamefont {Lorenz}\ \emph {et~al.}(2020)\citenamefont {Lorenz},
  \citenamefont {Erickson}, \citenamefont {Riesner}, \citenamefont {Gamelin},
  \citenamefont {Fainblat},\ and\ \citenamefont {Bacher}}]{lorenz2020directed}%
  \BibitemOpen
  \bibfield  {author} {\bibinfo {author} {\bibfnamefont {S.}~\bibnamefont
  {Lorenz}}, \bibinfo {author} {\bibfnamefont {C.~S.}\ \bibnamefont
  {Erickson}}, \bibinfo {author} {\bibfnamefont {M.}~\bibnamefont {Riesner}},
  \bibinfo {author} {\bibfnamefont {D.~R.}\ \bibnamefont {Gamelin}}, \bibinfo
  {author} {\bibfnamefont {R.}~\bibnamefont {Fainblat}},\ and\ \bibinfo
  {author} {\bibfnamefont {G.}~\bibnamefont {Bacher}},\ }\href@noop {}
  {\bibfield  {journal} {\bibinfo  {journal} {Nano Letters}\ }\textbf {\bibinfo
  {volume} {20}},\ \bibinfo {pages} {1896} (\bibinfo {year}
  {2020})}\BibitemShut {NoStop}%
\bibitem [{\citenamefont {Janet~Sherly}\ and\ \citenamefont
  {Nithiananthi}(2021)}]{janet2021diluted}%
  \BibitemOpen
  \bibfield  {author} {\bibinfo {author} {\bibfnamefont {I.}~\bibnamefont
  {Janet~Sherly}}\ and\ \bibinfo {author} {\bibfnamefont {P.}~\bibnamefont
  {Nithiananthi}},\ }\href@noop {} {\bibfield  {journal} {\bibinfo  {journal}
  {The European Physical Journal Plus}\ }\textbf {\bibinfo {volume} {136}},\
  \bibinfo {pages} {1} (\bibinfo {year} {2021})}\BibitemShut {NoStop}%
\bibitem [{\citenamefont {Sherly}\ and\ \citenamefont
  {Nithiananthi}(2021)}]{sherly2021tuning}%
  \BibitemOpen
  \bibfield  {author} {\bibinfo {author} {\bibfnamefont {I.~J.}\ \bibnamefont
  {Sherly}}\ and\ \bibinfo {author} {\bibfnamefont {P.}~\bibnamefont
  {Nithiananthi}},\ }\href@noop {} {\bibfield  {journal} {\bibinfo  {journal}
  {Physica B: Condensed Matter}\ }\textbf {\bibinfo {volume} {600}},\ \bibinfo
  {pages} {412615} (\bibinfo {year} {2021})}\BibitemShut {NoStop}%
\bibitem [{\citenamefont {Kalpana}\ and\ \citenamefont
  {Jayakumar}(2019)}]{kalpana2019magnetic}%
  \BibitemOpen
  \bibfield  {author} {\bibinfo {author} {\bibfnamefont {P.}~\bibnamefont
  {Kalpana}}\ and\ \bibinfo {author} {\bibfnamefont {K.}~\bibnamefont
  {Jayakumar}},\ }\href@noop {} {\bibfield  {journal} {\bibinfo  {journal}
  {Physica Scripta}\ }\textbf {\bibinfo {volume} {94}},\ \bibinfo {pages}
  {105817} (\bibinfo {year} {2019})}\BibitemShut {NoStop}%
\bibitem [{\citenamefont {Jayam}\ and\ \citenamefont
  {Navaneethakrishnan}(2002)}]{jayam2002optical}%
  \BibitemOpen
  \bibfield  {author} {\bibinfo {author} {\bibfnamefont {S.~G.}\ \bibnamefont
  {Jayam}}\ and\ \bibinfo {author} {\bibfnamefont {K.}~\bibnamefont
  {Navaneethakrishnan}},\ }\href@noop {} {\bibfield  {journal} {\bibinfo
  {journal} {International Journal of Modern Physics B}\ }\textbf {\bibinfo
  {volume} {16}},\ \bibinfo {pages} {3737} (\bibinfo {year}
  {2002})}\BibitemShut {NoStop}%
\bibitem [{\citenamefont {Kavokin}\ \emph {et~al.}(1999)\citenamefont
  {Kavokin}, \citenamefont {Merkulov}, \citenamefont {Yakovlev}, \citenamefont
  {Ossau},\ and\ \citenamefont {Landwehr}}]{kavokin1999exciton}%
  \BibitemOpen
  \bibfield  {author} {\bibinfo {author} {\bibfnamefont {K.}~\bibnamefont
  {Kavokin}}, \bibinfo {author} {\bibfnamefont {I.}~\bibnamefont {Merkulov}},
  \bibinfo {author} {\bibfnamefont {D.}~\bibnamefont {Yakovlev}}, \bibinfo
  {author} {\bibfnamefont {W.}~\bibnamefont {Ossau}},\ and\ \bibinfo {author}
  {\bibfnamefont {G.}~\bibnamefont {Landwehr}},\ }\href@noop {} {\bibfield
  {journal} {\bibinfo  {journal} {Physical Review B}\ }\textbf {\bibinfo
  {volume} {60}},\ \bibinfo {pages} {16499} (\bibinfo {year}
  {1999})}\BibitemShut {NoStop}%
\bibitem [{\citenamefont {Yakovlev}\ and\ \citenamefont
  {Ossau}(2010)}]{yakovlev2010magnetic}%
  \BibitemOpen
  \bibfield  {author} {\bibinfo {author} {\bibfnamefont {D.~R.}\ \bibnamefont
  {Yakovlev}}\ and\ \bibinfo {author} {\bibfnamefont {W.}~\bibnamefont
  {Ossau}},\ }in\ \href@noop {} {\emph {\bibinfo {booktitle} {Introduction to
  the Physics of Diluted Magnetic Semiconductors}}}\ (\bibinfo  {publisher}
  {Springer},\ \bibinfo {year} {2010})\ pp.\ \bibinfo {pages}
  {221--262}\BibitemShut {NoStop}%
\bibitem [{\citenamefont {Yakovlev}\ \emph {et~al.}(1992)\citenamefont
  {Yakovlev}, \citenamefont {Ossau}, \citenamefont {Landwehr}, \citenamefont
  {Bicknell-Tassius}, \citenamefont {Waag}, \citenamefont {Schmeusser},\ and\
  \citenamefont {Uraltsev}}]{yakovlev1992two}%
  \BibitemOpen
  \bibfield  {author} {\bibinfo {author} {\bibfnamefont {D.}~\bibnamefont
  {Yakovlev}}, \bibinfo {author} {\bibfnamefont {W.}~\bibnamefont {Ossau}},
  \bibinfo {author} {\bibfnamefont {G.}~\bibnamefont {Landwehr}}, \bibinfo
  {author} {\bibfnamefont {R.}~\bibnamefont {Bicknell-Tassius}}, \bibinfo
  {author} {\bibfnamefont {A.}~\bibnamefont {Waag}}, \bibinfo {author}
  {\bibfnamefont {S.}~\bibnamefont {Schmeusser}},\ and\ \bibinfo {author}
  {\bibfnamefont {I.}~\bibnamefont {Uraltsev}},\ }\href@noop {} {\bibfield
  {journal} {\bibinfo  {journal} {Solid state communications}\ }\textbf
  {\bibinfo {volume} {82}},\ \bibinfo {pages} {29} (\bibinfo {year}
  {1992})}\BibitemShut {NoStop}%
\bibitem [{\citenamefont {Lee}\ \emph {et~al.}(1996)\citenamefont {Lee},
  \citenamefont {Dobrowolska}, \citenamefont {Furdyna}, \citenamefont {Luo},\
  and\ \citenamefont {Ram-Mohan}}]{lee1996magneto}%
  \BibitemOpen
  \bibfield  {author} {\bibinfo {author} {\bibfnamefont {S.}~\bibnamefont
  {Lee}}, \bibinfo {author} {\bibfnamefont {M.}~\bibnamefont {Dobrowolska}},
  \bibinfo {author} {\bibfnamefont {J.}~\bibnamefont {Furdyna}}, \bibinfo
  {author} {\bibfnamefont {H.}~\bibnamefont {Luo}},\ and\ \bibinfo {author}
  {\bibfnamefont {L.}~\bibnamefont {Ram-Mohan}},\ }\href@noop {} {\bibfield
  {journal} {\bibinfo  {journal} {Physical Review B}\ }\textbf {\bibinfo
  {volume} {54}},\ \bibinfo {pages} {16939} (\bibinfo {year}
  {1996})}\BibitemShut {NoStop}%
\bibitem [{\citenamefont {Jain}(1991)}]{jain1991diluted}%
  \BibitemOpen
  \bibfield  {author} {\bibinfo {author} {\bibfnamefont {M.~K.}\ \bibnamefont
  {Jain}},\ }\href@noop {} {\emph {\bibinfo {title} {Diluted magnetic
  semiconductors}}}\ (\bibinfo  {publisher} {World Scientific},\ \bibinfo
  {year} {1991})\BibitemShut {NoStop}%
\bibitem [{\citenamefont {Kochereskho}\ \emph {et~al.}(1994)\citenamefont
  {Kochereskho}, \citenamefont {Merkulov}, \citenamefont {Pozina},
  \citenamefont {Uraltsev}, \citenamefont {Yakovlev}, \citenamefont {Ossau},
  \citenamefont {Waag},\ and\ \citenamefont {Landwehr}}]{kochereskho1994light}%
  \BibitemOpen
  \bibfield  {author} {\bibinfo {author} {\bibfnamefont {V.}~\bibnamefont
  {Kochereskho}}, \bibinfo {author} {\bibfnamefont {I.}~\bibnamefont
  {Merkulov}}, \bibinfo {author} {\bibfnamefont {G.}~\bibnamefont {Pozina}},
  \bibinfo {author} {\bibfnamefont {I.}~\bibnamefont {Uraltsev}}, \bibinfo
  {author} {\bibfnamefont {D.}~\bibnamefont {Yakovlev}}, \bibinfo {author}
  {\bibfnamefont {W.}~\bibnamefont {Ossau}}, \bibinfo {author} {\bibfnamefont
  {A.}~\bibnamefont {Waag}},\ and\ \bibinfo {author} {\bibfnamefont
  {G.}~\bibnamefont {Landwehr}},\ }\href@noop {} {\bibfield  {journal}
  {\bibinfo  {journal} {Solid-state electronics}\ }\textbf {\bibinfo {volume}
  {37}},\ \bibinfo {pages} {1081} (\bibinfo {year} {1994})}\BibitemShut
  {NoStop}%
\end{thebibliography}%

\end{document}